\documentclass{article}

\usepackage{arxiv}
\usepackage{graphicx,subfigure,amsmath,bbm}
\usepackage{moreverb}
\usepackage{rotating}
\usepackage{accents}
\usepackage[linesnumbered,ruled]{algorithm2e}
\usepackage{bm}

\usepackage[utf8]{inputenc}
\usepackage[english]{babel}
 \newtheorem{theorem}{Theorem}[section]
\newtheorem{corollary}{Corollary}[theorem]

\newtheorem{prop}{Proposition}

\usepackage[utf8]{inputenc} 
\usepackage[T1]{fontenc}    
\usepackage{hyperref}       
\usepackage{url}            
\usepackage{booktabs}       
\usepackage{amsfonts}       
\usepackage{nicefrac}       
\usepackage{microtype}      
\usepackage{lipsum}

\title{Logistic Box-Cox Regression to Assess the Shape and Median Effect under Uncertainty about Model Specification}

\author{
  Li~Xing\\
   Mathematics and Statistics\\ 
   University of Victoria \\
   Victoria, BC, Canada\\
   \And
   Xuekui~Zhang\\
   Mathematics and Statistics\\ 
   University of Victoria \\
   Victoria, BC, Canada\\
   \And
   Igor~Burstyn\\
   Department of Environmental and Occupational Health\\ 
   School of Public Health\\
   Drexel University\\
   Philadelphia, PA, USA\\
   \And
   Paul~Gustafson \thanks{Correspondence to Paul Gustafson, Department of Statistics, University of British Columbia, 3182 Earth Sciences Building, 2207 Main Mall, Vancouver, BC, Canada, V6T 1Z4. Email:   gustaf@stat.ubc.ca} \\
  Department of Statistics\\
  University of British Columbia\\
  Vancouver, BC, Canada\\
}

\begin{document}
\maketitle

\begin{abstract}
The shape of the relationship between a continuous exposure variable and a binary disease variable is often central to epidemiologic investigations. This paper investigates a number of issues surrounding inference and the shape of the relationship. Presuming that the relationship can be expressed in terms of regression coefficients and a shape parameter, we investigate how well the shape can be inferred in settings which might typify epidemiologic investigations and risk assessment. We also consider a suitable definition of the median effect of exposure, and investigate how precisely this can be inferred. This is done both in the case of using a model acknowledging uncertainty about the shape parameter and in the case of ignoring this uncertainty and using a two-step method, where in step one we transform the predictor and in step two we fit a simple linear model with transformed predictor. All these investigations require a family of exposure-disease relationships indexed by a shape parameter. For this purpose, we employ a family based on the Box-Cox transformation.
\end{abstract}


\section{Introduction}

Epidemiologists are often confronted with skewed distribution of exposure or dose-metrics (such as cumulative exposure) that is suspected to be related in a non-linear fashion with commonly employed functions of risk of a health outcome, such as is afforded by logistic regression. For instance, there may be saturation and threshold effects, as well as reversals of direction of association at different doses (e.g. drinking and heart health reported by Doll et al \cite{Doll911}). Therefore, the underlying assumption in the logistic regression about the linearity between the log-odds of disease and exposure may not be valid. As a remedy, researchers transform the exposure measurements using a logarithmic or square-root function and then plug the transformed measurements into a logistic model as the predictor. This data transformation step before model-fitting aims to make the relationship between the log-odds and the transformed exposure closer to linear. However, such two-step approach ignores the uncertainty in the nonlinear association by enforcing a logarithm or square root function, which lacks a theoretical justification for the choice of transformation function. Therefore, we build a parsimonious one-step model for two purposes. First we estimate a shape parameter in the model based on the maximum likelihood (ML) estimation from the data and this shape parameter shows the most likely nonlinear association type. Second the estimated shape parameter is an optimal transformation, which, in practice, provides the theoretical justification for the type of transformation for those who prefer the two step approach. Our discussion focuses on the general risk model for the association between a binary disease outcome, Y , and a continuous exposure variable, X.  Assume $X \sim \mbox{LN}(\mu, \sigma^{2})$ as is often realistic for environmental exposures \cite{KOCH1966276}. For the $i$-th subject, we have
\begin{equation} \label{eq:truemodel}
 \mbox{log}\left( \frac{p_{i}}{1-p_{i}} \right ) = \beta_{0} + \beta_{1} x_{i}^{(\lambda)}, 
\end{equation}
where $p_{i} = \mbox{E}( Y_{i} | X_{i} = x_{i} )$, $x_{i}^{(\lambda)} = \left(x_{i}^{\lambda} - 1\right)/\lambda$  for $\lambda > 0$, $x_{i}^{(\lambda)} = \mbox{log}(x_{i})$ for $\lambda = 0$, and $\lambda \geq 0$. The $x^{(\lambda)}$ function is a Box-Cox transformation \cite{Box1964}. Statistical models involving the Box-Cox transformation are discussed extensively in literature. In linear regression, due to requirement of normality of residuals, the Box-Cox transformation is often employed on the outcome variable \cite{HINKLEY:1984jo, CARROLL:1981bk, HINKLEY:1976dt, Box1964}. In both linear and logistic regressions, in order to satisfy linearity requirement between log-odds and predictors, the Box-Cox transformation is suggested for predictors \cite{Siqueira:1999ik, Kay:1987fy, Egger1979}.  We emphasize two desirable properties of the Box-Cox function: the continuity at $\lambda = 0$ and ability to accommodate several familiar transformations ( i.e., the logarithm function at $\lambda = 0$, the linear function at $\lambda = 1$, the square-root function at $\lambda = 0.5$, and the square function at $\lambda = 2$).

As a nonlinear model, the gradient of the log-odds of the logistic Box-Cox model is no longer constant. We are interested in this gradient with respect to $X^{(q)}$ for some choice of $q$. Particularly, we define  
\begin{equation}
Q_{q}=\frac{d(\mbox{logit}(\mbox{E}( Y | X ) ))}{dX^{(q)}}= \beta_{1} X^{\lambda - q}.
\end{equation} 
The quantity $Q_{q}$ represents the instantaneous effect of the predictor on the $X^{(q)}$ scale. In the Box-Cox model, if we can correctly specify $q = \lambda$, $Q_{q} (= \beta_{1})$ represents a constant effect on the $X^{(\lambda)}$ scale. For $q \ne \lambda$, the value of $Q_{q}$ changes over $X$ representing a non-constant effect on $X^{(q)}$. More specifically, $Q_{q}/\beta_{1}$ follows $\mbox{LN}((\lambda-q)\mu, (\lambda-q)^{2}\sigma^{2})$. Gelman and Pardoe \cite{GelmanPardoe2007} suggested averaging the effect of a predictor over the population distribution of predictors. Examples are shown in linear regression models \cite{Liu2008} and in the survival analysis context \cite{Gustafson2007}. We adapt their definitions to the logistic regression context to arrive at a summary of $Q_{q}$.  We define the average effect, $\Delta_{q}$, and the median effect, $\Delta^{*}_{q}$, as summary measurements of the effect of the predictor $X$ on the $X^{(q)}$ scale in the following.
\begin{equation}
\Delta_{q} =  \mbox{E}(Q_{q}) = \beta_{1} \mbox{exp}\left((\lambda-q)\mu+\frac{(\lambda-q)^2\sigma^{2}}{2}\right), 
\end{equation}
and 
\begin{equation}
\Delta^{*}_{q} =  \mbox{Median}(Q_{q}) = \beta_{1} \mbox{exp}\left((\lambda-q)\mu\right).
\end{equation}
Because median is more robust than mean for a long-tailed distribution, going forward we adopt the median effect, $\Delta^{*}_{q}$, as the representative of the overall gradient of the log-odds. 

In Section 2, we provide two propositions on the MLE of the logistic Box-Cox model and propose an optimization algorithm to obtain the MLE. In Section 3, we discuss the misspecified logistic linear model and define a quantity to measure the distance between the median effect and the slope coefficient estimated from the misspecified model. In Section 4, we design and conduct simulation studies to evaluate the accuracy of the parameter estimates of the logistic Box-Cox model based on the quasi-Newton method, to compare the median effect and its approximation from a simple linear model with transformed predictor, and to calculate the asymptotic standard deviations of the model parameter estimates as well as that of the estimated median effect. In Section 5, we apply our model to a real data set and compare this model with three two-step approaches. In Section 6, we summarize our results and draw conclusions.           

\section{The Logistic Box-Cox Model}

In this section, we prove that the log-likelihood function of the logistic Box-Cox model is strictly concave. So to obtain MLE, we only need to find the root of the score function. Based on this property and optimization methods for this model in the literature, we use the quasi-Newton algorithm to compute the MLE. In addition, we use numerical methods to approximate the asymptotic variance of the MLE, which help us understand the precision of the parameter estimates under large samples and also help design our future experiments.   

\begin{prop}
The Hessian matrix of the log-likelihood function of the logistic Box-Cox model is negative definite for any interior point in the three dimensional space $(-\infty, +\infty) \times (-\infty, +\infty) \times [0, +\infty)$. 
\end{prop}
\begin{corollary}
The log-likelihood function is strictly concave and, therefore, any root of the score function is the unique global maximum of the likelihood function. 
\end{corollary}
The proof of the Proposition 1 is given in the appendix and the proof of the corollary is trivial. In the literature, there are two kinds of optimization methods for this model. Egger \cite{Egger1979} mentioned the difficulty of convergence for the Newton-Raphson method in practice and  suggested using the profile likelihood (PL) method, where we do a grid search on the shape parameter $\lambda$, use iteratively re-weighted least squares to estimate the regression coefficients, $\beta_{0}$ and $\beta_{1}$, given each fixed $\lambda$, and choose the set of estimates based on ML. Guerrero et al \cite{Guerrero1982} suggested a quasi-Newton method to estimate the parameters of the logistic Box-Cox model. Different from the Newton-Raphson method, in the quasi-Newton method, we replace the inverse of the Hessian matrix by an approximation in each iteration. This can reduce the numerical non-stability in getting the inverse of a matrix. As the log-likelihood has such nice properties, we choose the quasi-Newton method, but use the PL method to obtain the initial points. Particularly, the quasi-Newton method that we employ is the Broyden-Fletcher-Goldfarb-Shanno (BFGS) optimization method (\cite{Fletcher:1970fs, BROYDEN:1970jk, 10.2307/2004873, 10.2307/2004840}), which has been written in a wrapper function in the r package, maxLik \cite{Henningsen:2010hd}.  

\begin{prop}
\[\mbox{Avar}(\hat{\lambda}) = \mbox{O}(\beta_{1}^{-2}).\]
\end{prop}
The proof of Proposition 2 is also given in the appendix. This proposition demonstrates that, under a weak association between the predictor and outcome variables (i.e. small value of $\beta_{1}$), in order to get a precise estimate of the shape parameter, we need a large sample size as we know that $\mbox{Var}(\hat{\lambda}) \approx \mbox{Avar}(\hat{\lambda})/n$. We calculate the asymptotic variance of the model parameters based on inverse of the Fisher information matrix through numerical methods, where details are in the appendix, and we calculate the asymptotic variance of the median effect, $\Delta^{*}_{q}$, based on the multivariate delta method listed below. 
\begin{eqnarray}
\nonumber \mbox{Avar}\left( \hat{\Delta}^{*}_{q} \right) &\approx & \left(\begin{array}{cc}
        \frac{\partial \Delta^{*}_{q}}{\partial \beta_{1}}, &
        \frac{\partial \Delta^{*}_{q}}{\partial \lambda}
               \end{array}\right) \left (\begin{array}{ll}
 \mbox{Avar}(\hat{\beta}_{1}),  & \mbox{Acov}(\hat{\beta}_{1}, \hat{\lambda}) \\
 \mbox{Acov}(\hat{\beta}_{1}, \hat{\lambda}), &  \mbox{Avar}(\hat{\lambda})  
\end{array}\right) \left(\begin{array}{c}
        \frac{\partial \Delta_{q}^{*}}{\partial \beta_{1}}\\
        \frac{\partial \Delta^{*}_{q}}{\partial \lambda}
               \end{array}\right)\\   
\label{eq:VarDelta} &=& \Delta^{*2}_{q} \left(\frac{\mbox{Avar}(\hat{\beta}_{1})}{\beta_{1}^{2}} + \frac{2\mu}{\beta_{1}}\mbox{Acov}(\hat{\beta}_{1}, \hat{\lambda}) + \mu^{2}\mbox{Avar}(\hat{\lambda})\right).   
\end{eqnarray}

\section{The Misspecified Logistic Linear Model} \label{sec:mismodel}
Assume that the true model is a logistic Box-Cox model. We are interested in the bias incurred if we fit a misspecified logistic linear model with a Box-Cox transformed $X$ as a predictor. In the misspecified model, the type of Box-Cox transformation is given beforehand, which means the shape parameter, $q$, is a fixed constant. We denote the transformed predictor as $W_{q}$, where   
\begin{equation}
W_{q} = X^{(q)} = \left\{ \begin{array}{cc}
 \frac{X^{q} - 1}{q} & \mbox{if } q > 0, \\
\mbox{log}(X)& \mbox{if }  q = 0.
\end{array} \right .   
\end{equation}
The misspecified model is written as below.
\begin{equation} \label{eq:mismodel}
\mbox{logit} \left(\mbox{Pr} \left (Y = 1 \left| W_{q} = w_{q} \right. \right)  \right) = \gamma_{0q} + \gamma_{1q} w_{q}.
\end{equation}
To obtain the large-sample limit of the estimated coefficients, $(\hat{\gamma}_{0q}, \hat{\gamma}_{1q})$, we need to solve the following equations:
\begin{equation} \label{eq:likeq}
 \bm{E} \left[ \left( \begin{array}{c}
  1\\
  W_{q}
  \end{array}\right)
    \left( \mbox{expit}(\beta_{0} + \beta_{1}X^{(\lambda)}) -
            \mbox{expit}(\gamma_{0q} + \gamma_{1q}W_{q}) 
    \right)
\right] =  \bm{0},
\end{equation} 
where $\mbox{expit}(\cdot) = \mbox{exp}(\cdot)/(1+\mbox{exp}(\cdot))$.
Due to the misspecified likelihood, the inverse of the Fisher Information matrix is no longer providing the asymptotic variances of the parameter estimates. Therefore, we use the sandwich type estimates \cite{White1982, freedman2006so} for the variance estimates, whereby
\begin{equation} \label{eq:AvarGamma}
\mbox{Avar}(\hat{\bm \gamma}_{q}) \approx J^{-1}_{1}(\hat{\bm \gamma}_{q})V_{1}(\hat{\bm \gamma}_{q})J^{-1}_{1}(\hat{\bm \gamma}_{q}),
\end{equation} 
where $J_{1} = \mbox{E}(H(l_{1}))$ with $l_{1}$ representing the likelihood function of the model (~\ref{eq:mismodel}) and $H(l_{1})$ representing the Hessian matrix, $V_{1} = \mbox{Var}(\nabla l_{1})$, and $\hat{\bm \gamma}_{q} = (\hat{\gamma}_{0q}, \hat{\gamma}_{1q})^{T}$ is the solution of (~\ref{eq:likeq}). More detailed mathematical work is provided in the appendix. 

\section{Simulation Studies}

In the simulation studies, our aims are three-fold: (1) evaluating the accuracy of the parameter estimates in the logistic Box-Cox model based on the BFGS method, (2) comparing the distance between the median effect from the underlying logistic Box-Cox model with its approximation, the large sample limit of the  estimate of the slope parameter from the misspecified linear model, and (3) calculating the asymptotic standard deviations of the model parameter estimates as well as that of the estimated median effect. To achieve these aims we design a factorial experiment, using factors whose levels reflect plausible contexts for epidemiologic investigations. 

\subsection{Simulation Design}
We choose four factors to control our simulation experiment, which are: 
\begin{enumerate}
\item the shape of exposure distribution;
\item the shape of exposure-disease relationship;
\item the disease rarity;
\item the strength of exposure-disease association.
\end{enumerate}

Table ~\ref{tab:factors} shows us the levels of each factor. First, without loss of generality, we fix the $95$-th percentile of the distribution of $X$ at $1$ and vary $\sigma$ to control the level of skewness of the distribution of $X$. Second, we vary the shape parameter, $\lambda$, as $0, 0.5, 1$ and $2$, which corresponds to log, square-root, linear and square functions respectively. Third, we use the probability of disease at the $5$-th percentile of the exposure to indicate the disease rarity, varying this as $\mbox{P}_{1} = 0.02$ and $\mbox{P}_{2} = 0.1$.  Fourth, we consider the ratio of the probability of the disease at $95$-th percentile of the exposure to the probability of the disease at $5$-th percentile, which is denoted as 
\begin{equation}
\mbox{R} = \frac{\mbox{Pr}(Y = 1| X \mbox{ is at } 95\mbox{-th percentile})}{\mbox{Pr}(Y = 1| X \mbox{ is at } 5\mbox{-th percentile})}.    
\end{equation}
We let $\mbox{R}_{1} = 1.1,  \mbox{R}_{2} = 2,$ and $\mbox{R}_{3} = 5$ to represent weak, medium and strong associations respectively.

\begin{figure}[t]
\centerline{\includegraphics[width=15cm, height=10cm]{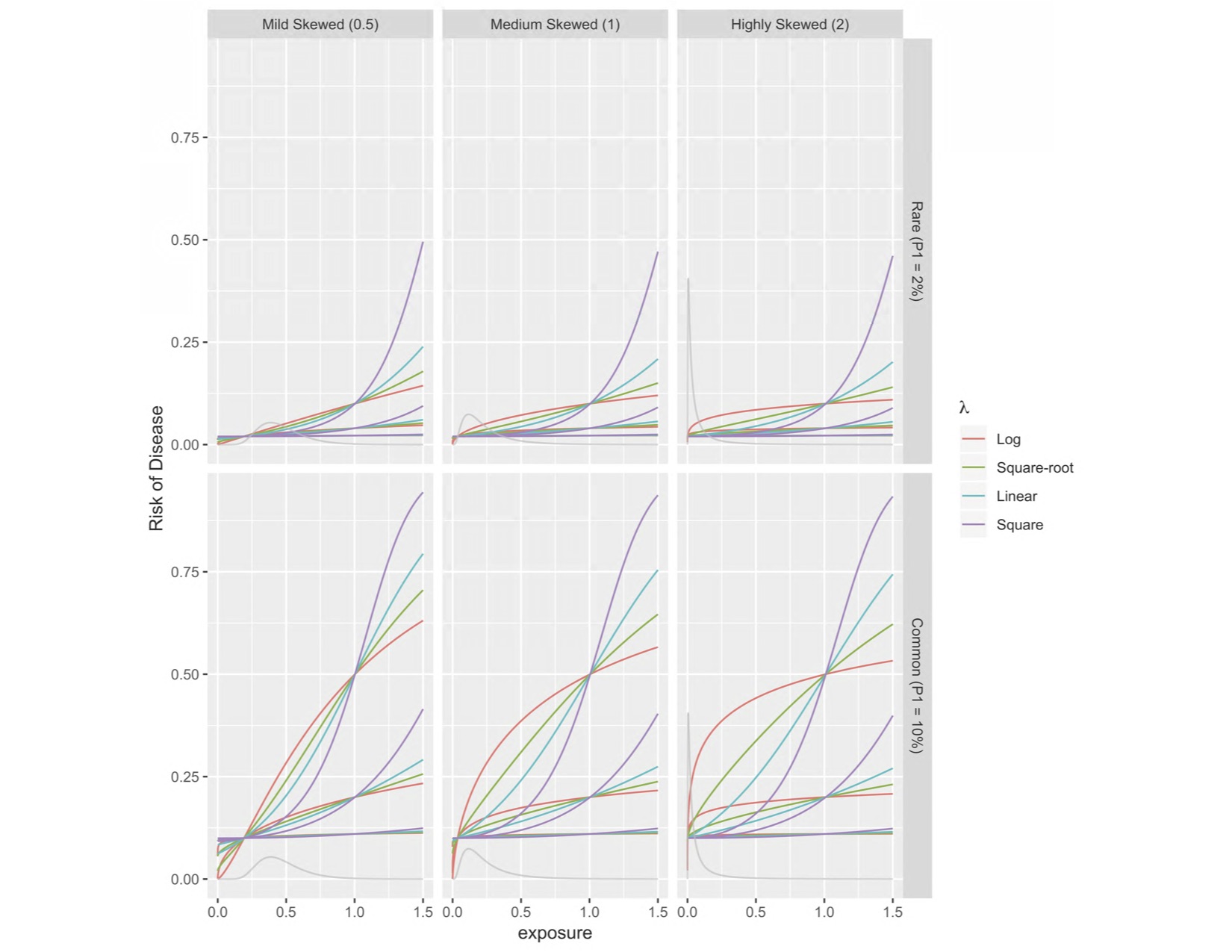}}
\caption{The risks of disease as a function of exposure in the selected settings. The distribution of the exposure for each panel is shown at the bottom of each panel by its density curve in grey.\label{fig1}}
\end{figure}

Figure ~\ref{fig1} shows the disease risk as a function of the exposure in the described 72 settings. In each panel, the distribution of exposure and the disease rarity is fixed. The risk functions vary with the shape parameters in the model and the risk ratios indicating the strength of the association. We can see that given the distribution of exposure, as the exposure-disease association becomes stronger, the risk differences between different shape parameters at the same exposure level become larger. Also, given the association, as the distribution becomes more skewed, the risk differences between different shape parameters at the same exposure level become larger. These indicate that the skewness and the strength of association may be related to the precision in estimating the shape parameter.  Also this figure illustrates the magnitude of the risk and the gradient of the log-odds with respect to $X$ under different experimental settings. This can help us understand the real data under the similar conditions and also guide our future experiments. 
 
\subsection{Simulation Results}
\subsubsection{Aim 1: Evaluation of the BFGS method}

In this simulation, under each setting, we generate $500$ data sets, for each of which we generate $5000$ $X$'s as $\mbox{LN}(\mu, \sigma^{2})$, and the corresponding $Y$'s from the Bernoulli distribution with probability $\mbox{P} =  \mbox{expit}(\beta_{0} + \beta_{1}X^{(\lambda)})$. For each data set, we apply the PL method firstly, and use the estimates of the PL method as the initial points for the BFGS method. 

\begin{figure}[t]
\centerline{\includegraphics[width=12cm, height=8cm]{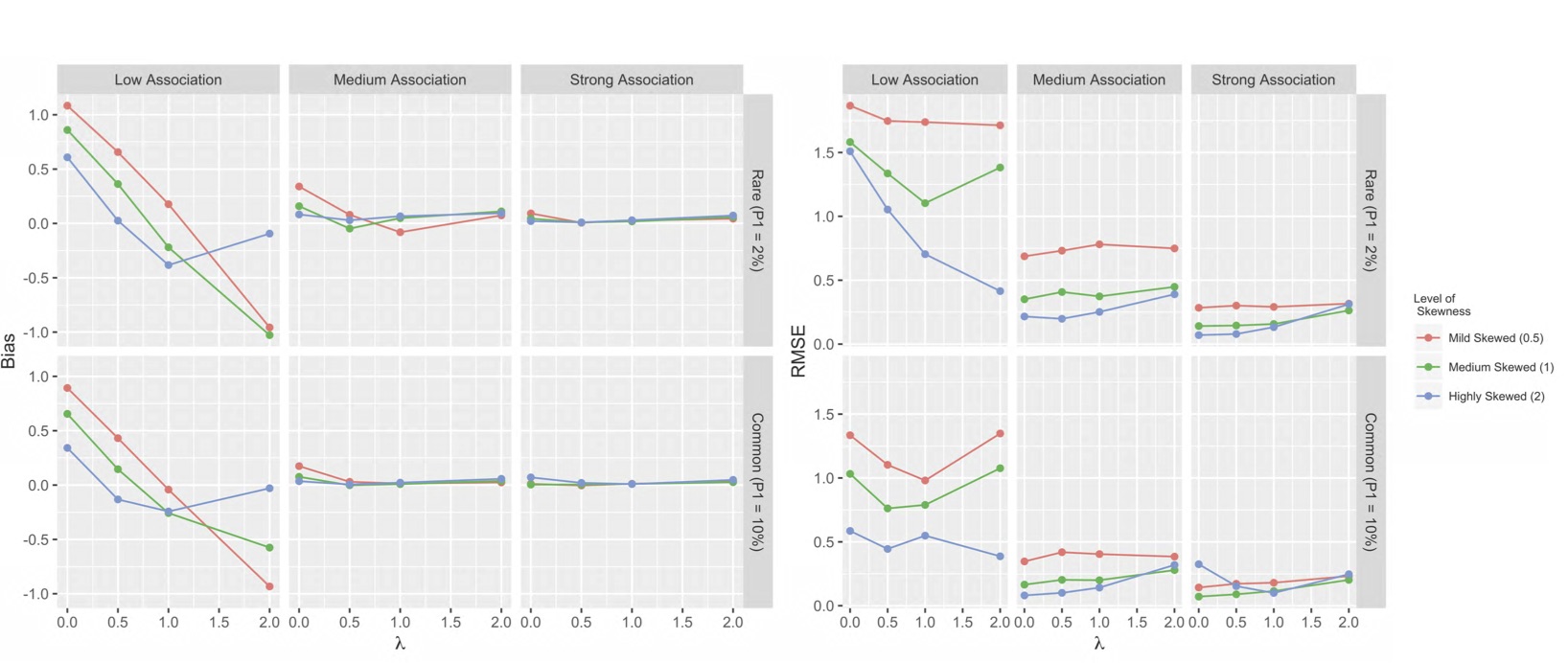}}
\caption{ The bias (in the left panel) and root mean squared error (RMSE) (in the right panel) of $\hat{\lambda}$ from the BFGS methods under different settings.}\label{fig:fig2}
\end{figure}
Figure ~\ref{fig:fig2} demonstrates that the ML estimation implemented with the BFGS algorithm provides fairly accurate estimation of $\lambda$ when the exposure variable and disease outcome have some degree of association. When their association is very weak, the bias and RMSE are considerably larger. However, with the medium level of association the bias is much smaller than $0.5$. This suggests we can easily distinguish a linear transformation from a square-root one or a square-root one from a log one. The stronger the association is the more accurate the estimates are. The results confirm that the BFGS method works well for our model fitting. The figures for bias and RMSE in estimating other parameters are provided in the appendix.        

\subsubsection{Aim 2: The Gradient Measurement of the Logistic Box-Cox Model and Its Estimate}
In the logistic Box-Cox model, we use the median effect, $\Delta^{*}_{q}$, to represent the gradient of the log-odds with respect to $X^{(q)}$ scale. We hope that if the sample is large enough, the estimate of the slope coefficient, $\hat{\gamma}_{1q}$ from the misspecified logistic linear model with $X^{(q)}$ as the predictor can be a good approximation of $\Delta^{*}_{q}$. Therefore, we define the absolute relative error (ARE) to measure the difference between the large sample limit, $\gamma_{1q}$, and $\Delta^{*}_{q}$.
That is, 
\begin{equation} 
\mbox{ARE} = \left |\frac{\gamma_{1q} - \Delta^{*}_{q} }{ \Delta^{*}_{q}} \right |,
\end{equation}
where $\gamma_{1q}$ is defined in the misspecified model (~\ref{eq:mismodel}). 
Under each of the $72$ settings, we fix the level of the other three factors and let the value of $\lambda$ vary from $0$ to $2$ with an increment of $0.25$. And for each $\lambda$, we generate $50,000$ of $X$'s from $\mbox{LN}(\mu, \sigma^2)$ and then we have the corresponding probability $\mbox{P} =  \mbox{expit}(\beta_{0} + \beta_{1}X^{(\lambda)})$. We vary $q$ from $0$ to $2$ with an increment of $0.25$. For each $q$, we approximate expectation of the functions in equation (~\ref{eq:likeq}) by their sample mean from samples of the $50,000$ of $X$'s and then solve the equations to get the limiting coefficient, $\gamma_{1q}$. As we have the true value of $\Delta^{*}_{q}$, we get the numerically approximated AREs as a function of $\lambda$ and $q$ under each setting. 

\begin{figure}[t]
\centerline{\includegraphics[width=10cm, height=8cm]{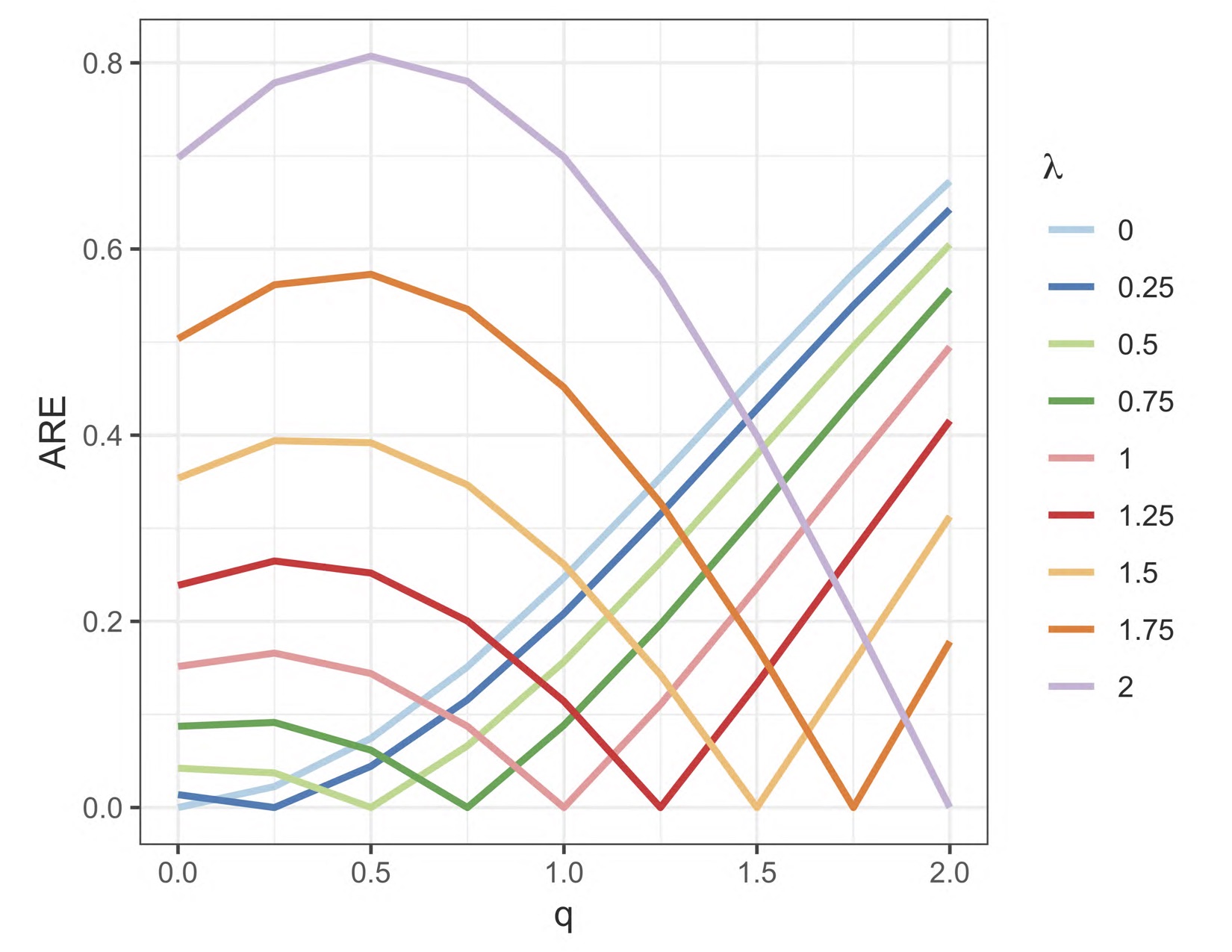}}
\caption{The estimated absolute relative error under the disease rarity $\mbox{P}_{1} = 0.02$, a weak association between exposure and disease $\mbox{R}_{1} = 1.1$, and a medium skewed distribution of exposure $X$ with $\sigma = 1$. \label{fig3}}
\end{figure}

Figure ~\ref{fig3} shows ARE as a function of $\lambda$ and $q$ under the settings with rare disease, weak association, and medium skewed distribution. The pattern is similar for all the settings. We can see that for each $\lambda$, when $q$ approaches $\lambda$ from the right side, ARE monotonically decreases and the rate of decrease is close to constant. When $q$ approaches $\lambda$ from the left side, ARE behaves like a quadratic function, with a maximum point between $0$ and $\lambda$. Therefore, to get smaller ARE, we suggest that it is safer to guess $q = 0$. Though the patterns are similar among all settings, the ARE inflates under conditions of strong association, common disease rarity, and more skewed distribution of $X$. 

\subsubsection{Aim 3: Calculation of the Asymptotic Standard Deviations}

Without loss of generality, we calculate the asymptotic standard deviation (ASD) of $\hat{\lambda}$ for a dataset with one observation based on numerical approximation of the inverse of the expected Fisher information matrix described in the appendix.  
\begin{figure}[t]
\centerline{\includegraphics[width=15cm, height=10cm]{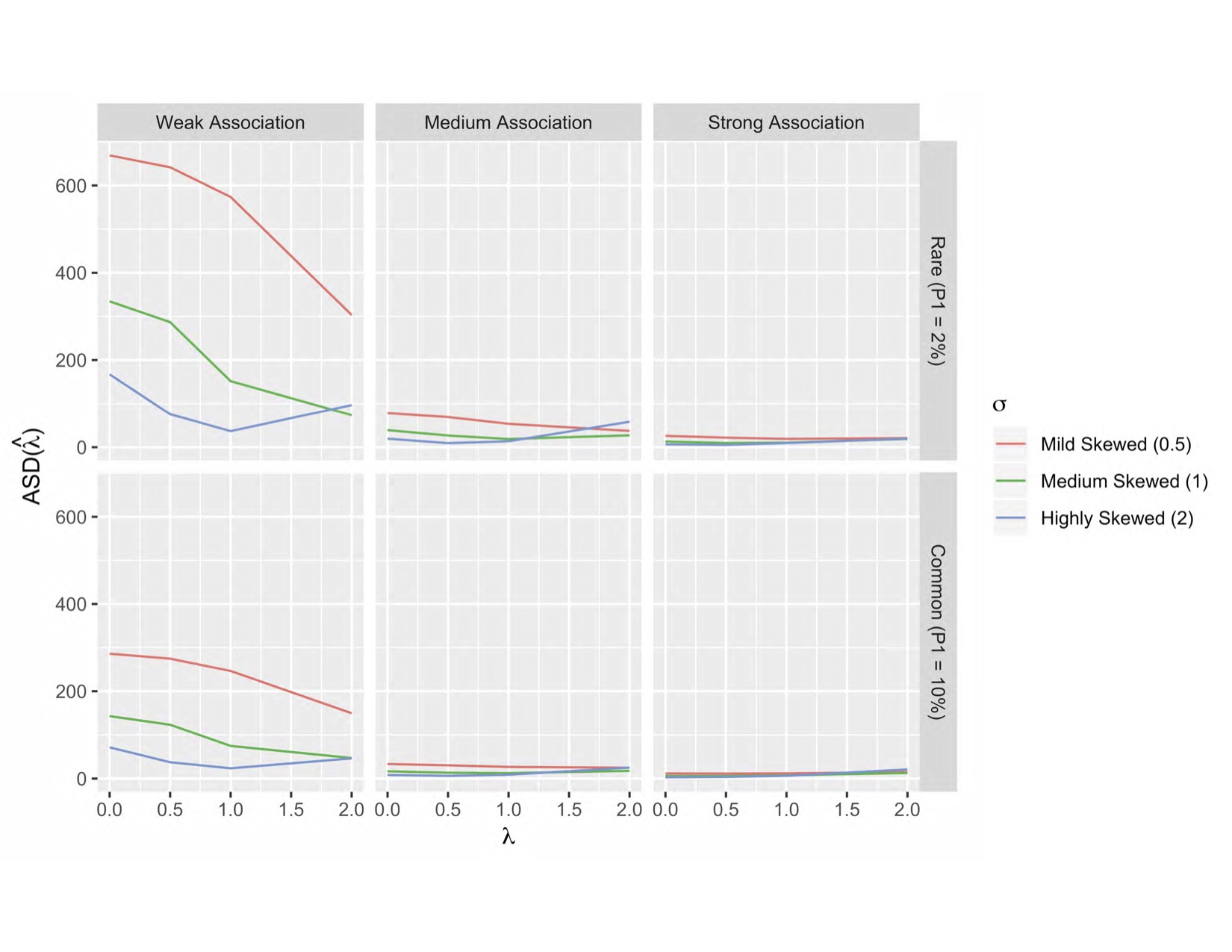}}
\caption{The asymptotic standard deviations of $\hat{\lambda}$ from the logistic Box-Cox model under different settings. \label{fig4}}
\end{figure}
Figure ~\ref{fig4} demonstrates that $\mbox{ASD}(\hat{\lambda})$ decreases when the association becomes stronger, when the disease becomes more common, or when the predictor is more skewed given other conditions do not change. Particularly, under a weak association, $\mbox{R}_{1}$, and a rare disease situation, $\mbox{P}_{1}$,  $\mbox{ASD}(\hat{\lambda}$)  is much larger than in the other situations, which indicates that when information is weak, it is harder to determine of the value of $\lambda$. The numerically approximated $\mbox{ASD}(\hat{\lambda})$ can help us design future studies. For example, if we would like to detect the difference of $0.5$ in the estimate of $\lambda$ in order to distinguish between a logarithm transformation and a square-root transformation, the standard error (SE) of $\hat{\lambda}$ should be less than $0.125$. We can achieve this by adding more samples. Under $\mbox{P}_{1}, \mbox{R}_{1}, \lambda = 0$ and the weakly skewed exposure, $\mbox{ASD}(\hat{\lambda}) \approx 700$ so that the sample size required to make $\mbox{SE}(\hat{\lambda}) = 0.125$ is equal to $(700/0.125)^2 = 31,360,000$. Therefore, any sample size larger than $31,360,000$ can provide us the power to distinguish the difference of $0.5$ in the estimate of $\lambda$ under the weakest condition, while this requirement decreases to less than one fourth of the big number when the condition changes to $\mbox{P}_{2}$ and others maintain the same. Note that in virtually all cases, it is not feasible to recruit around 30 millions participants in a study. To achieve this precision, the least requirement of the sample size among all of the settings of consideration is only $(2.914/0.125)^2 = 544$, which is under $\mbox{P}_{2}, \mbox{R}_{2}, \lambda = 0,$ and $\sigma = 2$. 

We also calculate $\mbox{ASD}(\hat{\gamma}_{1q})$ for a single-observation dataset based on the sandwich method for the misspecified likelihood and numerical methods, as discussed in Section ~\ref{sec:mismodel} and the appendix. In addition, we vary $q = 0, 0.5, 1,$ and $2$ to understand the difference across $q$.
\begin{figure}[t]
\centerline{\includegraphics[width=15cm, height=10cm]{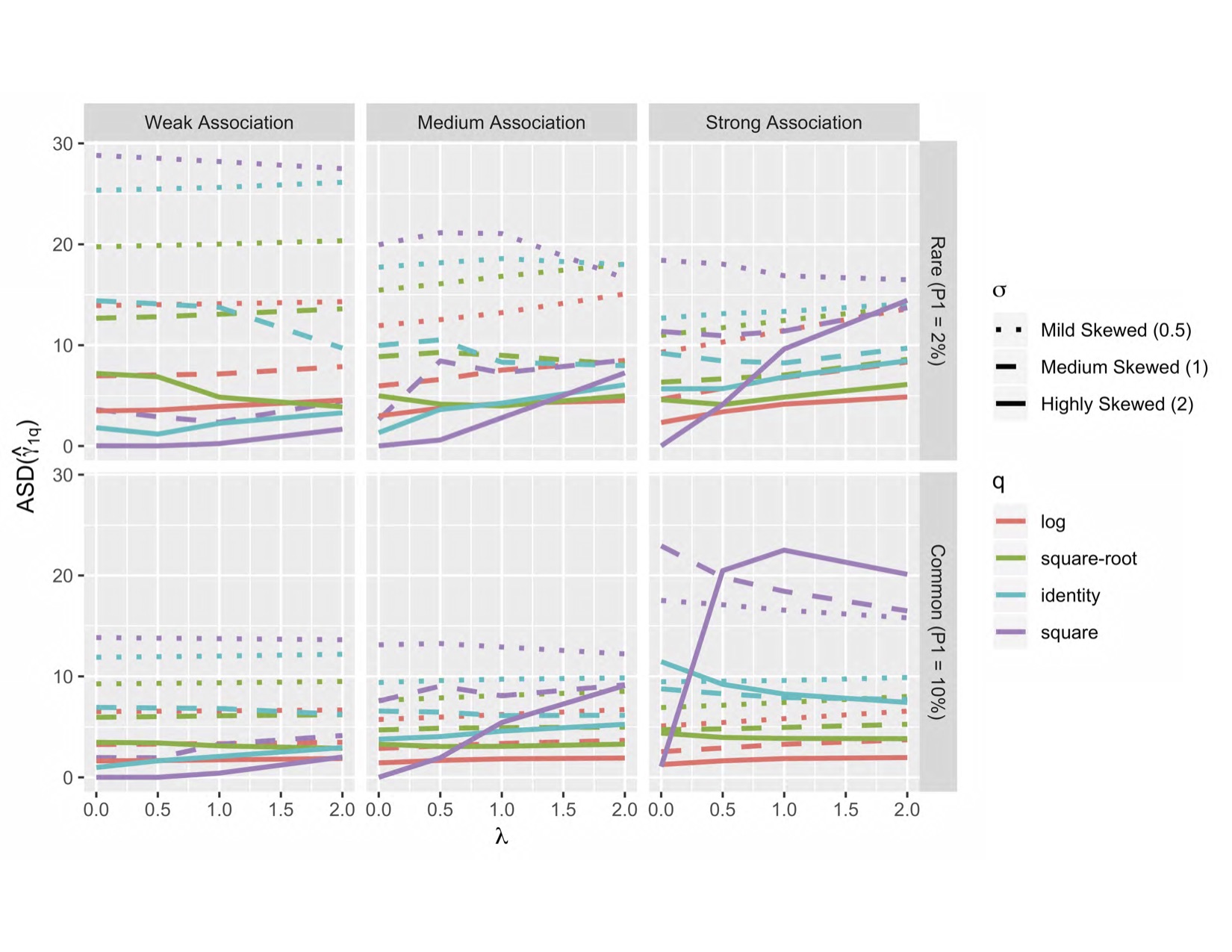}}
\caption{The asymptotic standard deviation of $\hat{\gamma}_{1q}$ from the misspecified logistic linear model under different settings.  \label{fig5}}
\end{figure}
Figure ~\ref{fig5} illustrates that $\mbox{ASD}(\hat{\gamma}_{1q})$ inflates under two extreme conditions. One is under weak association, rare disease, mild skewed distribution of $X$ and $q = 2$. The other is under strong association, common disease, and $q = 2$. When there is weak  association, rare disease and mild skewed condition, we can not get a precise estimate of the slope based on the misspecified linear model on any of the examined scales of $X$. On the other side, when there is strong association and common disease, we can not get a precise estimate of the slope if we enforce a linear pattern on a square scale. In general, $\mbox{ASD}(\hat{\gamma}_{1q})$ with $q=0$ is relatively low under all situations, though the precision worsen slightly when $\lambda$ is further from $0$. This implies that when there is little information, a logarithm transformation is a safer guess.      
 
Finally, we calculate $\hat{\Delta}^{*}_{q}$ based on the multivariate delta method.  
\begin{figure}[t]
\centerline{\includegraphics[width=15cm, height=10cm]{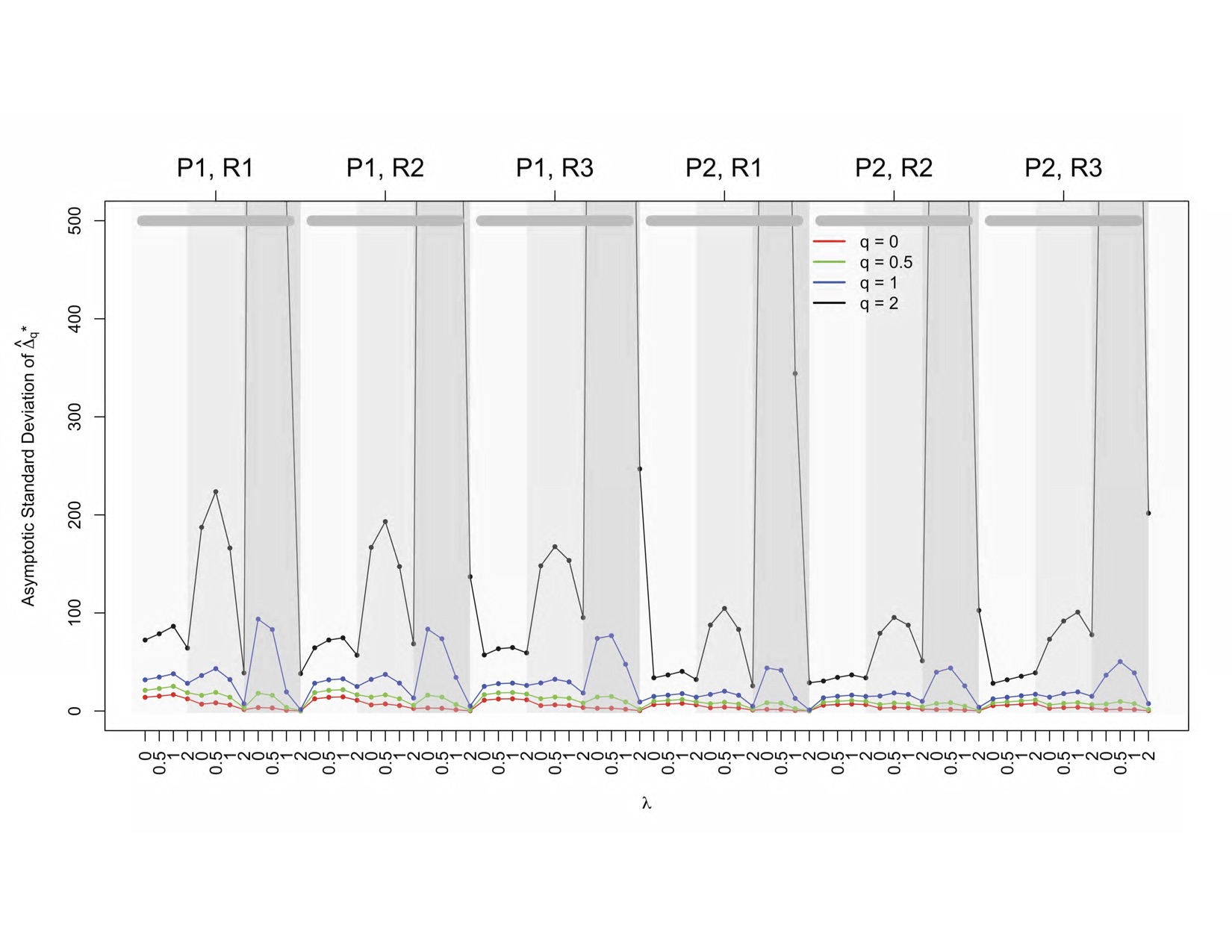}}
\caption{The asymptotic standard deviation of $\hat{\Delta}^{*}_{q}$ from the logistic Box-Cox model under different settings. Our setting are (1) disease rarity $\mbox{P}_{1} = 0.02$ and $\mbox{P}_{2} = 0.1$; (2) association between exposure and disease $\mbox{R}_{1} = 1.1, \mbox{R}_{2} = 2, $ and $\mbox{R}_{3} = 5$; (3) distribution of exposure $X$ $\sigma = 0.5$ (shaded light gray area), $\sigma = 1$ (shaded gray area) and $\sigma = 2$ (shaded dark gray area); (4) the shape of the relationship $\lambda = 0$ (log), $\lambda = 0.5$ (square-root), $\lambda = 1$ (linear) and $\lambda = 2$ (square). \label{fig6}}
\end{figure}
Figure ~\ref{fig6} illustrates that under all experimental settings, $\mbox{ASD}(\hat{\Delta}^{*}_{q})$ is monotonically increasing as a function of $q$. This makes sense since when $q$ becomes smaller, $\mbox{ASD}(\hat{\Delta}^{*}_{q})$ shows the gradient at a slower changing scale. Therefore, $\mbox{ASD}(\hat{\Delta}^{*}_{0})$ is always the smallest for each setting, which indicates precise estimation of the median effect on the log scale.      

\section{Application}

We analyze data from the National Health and Nutrition Examination Survey (NHANES) $2009$-$2010$, which involves $9,781$ adults aged $40$ years and above with measurements of both total blood mercury and depression. The exposure variable, $X$, is the total blood mercury in microgram per liter (ug/L), and the binary outcome, $Y$, is dichotomized from the score of the Patient Health Questionnaire-$9$ (PHQ-$9$) with $0$ indicating no depression (PHQ-$9$ score $\leq 9$) and $1$ indicating depression (PHQ-$9$ score $\geq 10$). 
\begin{figure}[t]
\centerline{\includegraphics[width=10cm, height=8cm]{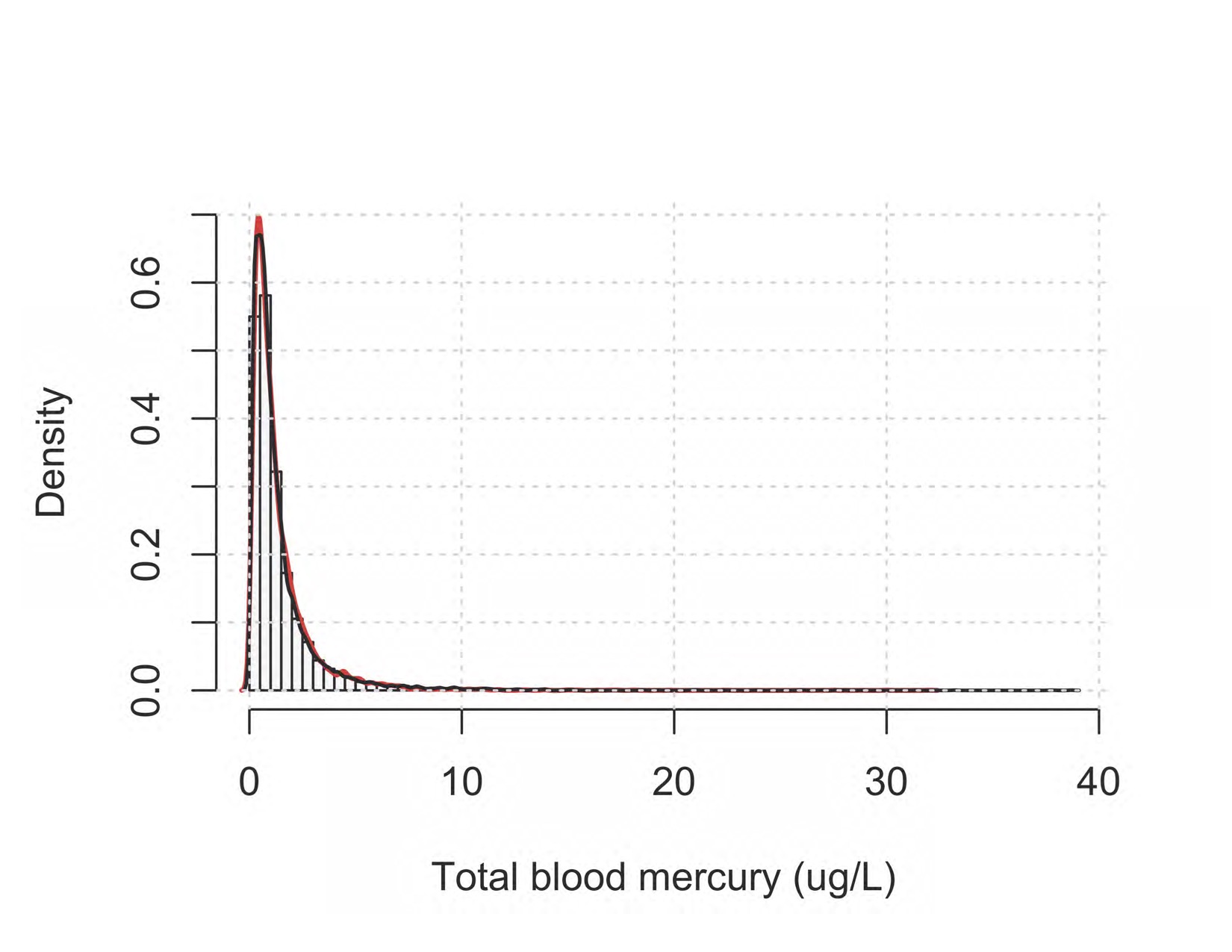}}
\caption{The histogram of the total blood mercury in microgram per liter. The black curve is the kernel density estimate
of the total blood mercury and the red curve is the probability density curve of log-normal$(-0.12, 0.93)$.\label{fig7}}
\end{figure}
Shown in Figure ~\ref{fig7}, the total blood mercury is right-skewed. And its distribution is approximated by the log-normal with $\hat{\mu} = -0.12$  and $\hat{\sigma} = 0.93$. We fit the logistic Box-Cox model relating the blood mercury level to prevalence of depression. The estimated parameters are $\hat{\beta}_{0} = -2.469$ (SE $0.046$),  $\hat{\beta}_{1} = -0.317$ (SE $0.046$), and $\hat{\lambda} =  0.392$ (SE $0.191$). Therefore, we can see that the estimated prevalence of depression at $ X = 0 $ is $\hat{\mbox{Pr}}(Y = 1 | X = 0) = \mbox{expit}(\hat{\beta}_{0} + \hat{\beta}_{1}X^{(\hat{\lambda})})|_{X=0}= 16\%$. And the instantaneous risk decline rate of prevalence of depression at certain level of the total blood mercury can be calculated. For instance, at $X=1$, this rate is   
\begin{equation*}
\left.\frac{\partial }{\partial X}\left[\mbox{Pr}(Y=1|X)\right]\right|_{X=1} = \left.\frac{\partial }{\partial X}\left[ \mbox{expit}(\beta_{0} + \beta_{1}X^{(\lambda)}) \right] \right|_{X=1}= \frac{\beta_{1}\mbox{exp}(\beta_{0})}{\left(1 + \mbox{exp}(\beta_{0})\right)^{2}} 
\end{equation*}
Plugging in the estimated coefficients, we get its estimate, $-0.022.$
The estimated median effect on $X$, $\hat{\Delta}^{*}_{1}$, is $-0.613$ with the $95\%$ point-wise confidence interval $[-0.978, -0.279]$, showing an overall negative association between the total blood mercury and depression. 
Next we would like to compare this fitted Box-Cox model with the misspecified linear model on $X^{(q)}$ scale for $q = 0, 0.5$ and $1$. 
In Table \ref{tab:modelXq}, we include the estimated slope coefficients and the Akaike information criterion (AIC) of the misspecified models, the corresponding estimated median effects from the logistic Box-Cox model, and the estimated ARE between the estimated median effect and the estimated slope coefficient from the data. When $q = 0.5$, we have the minimum AIC, which suggests the square-root model is the best among the three misspecified models. On the other hand, if we look at ARE, the smallest ARE occurs between the log model and the logistic Box-Cox model.  
\begin{figure}[t]
\centerline{\includegraphics[width=10cm, height=8cm]{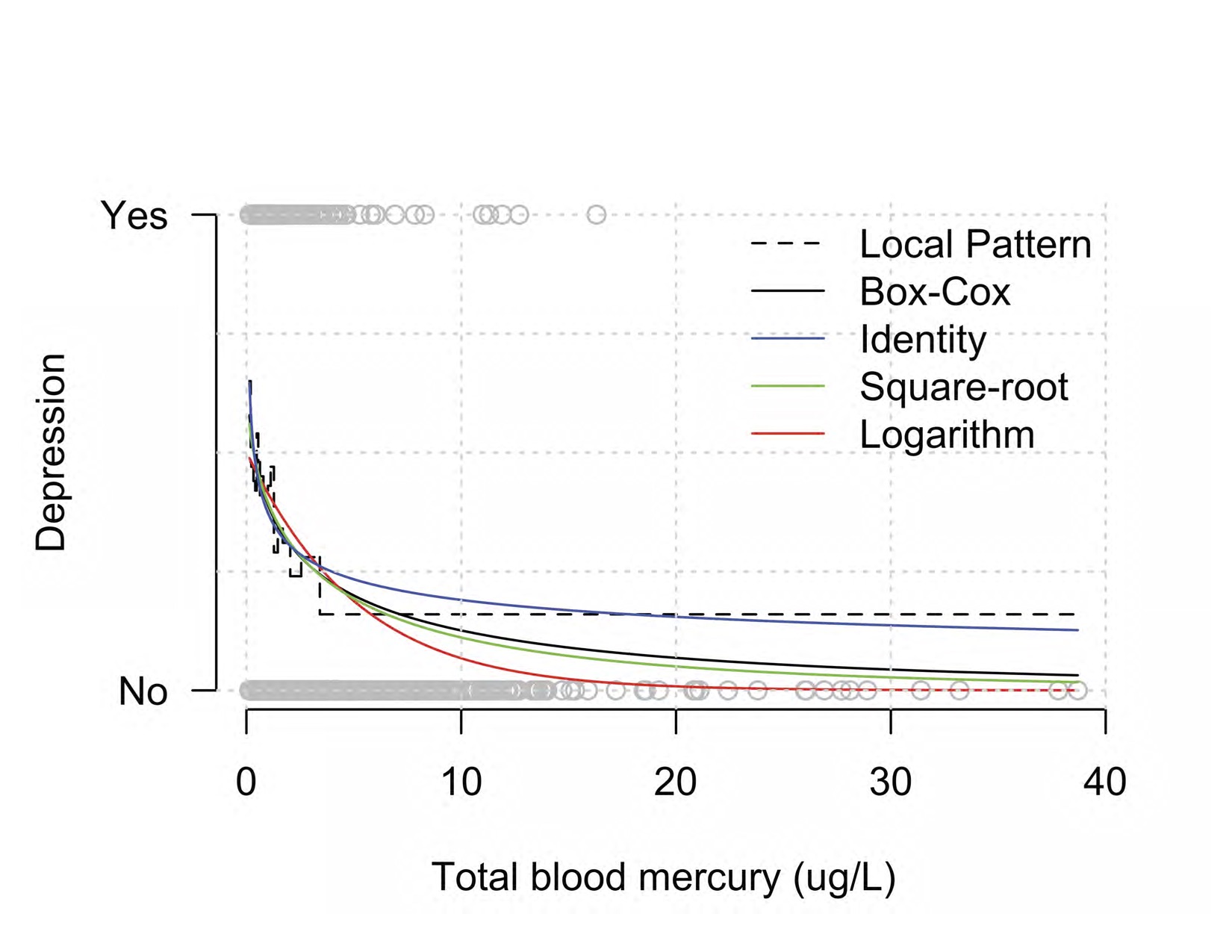}}
\caption{The local pattern of risks and four fitted models. The gray circles represent the raw data. The black dashed segmented lines represent the rate of disease at the adjacent intervals.\label{fig8}}
\end{figure}
To illustrate the local pattern of the relationship between the total blood mercury level and depression, we use a three-step procedure. First, we sort the data based on the blood mercury level from small to large. Second, we bin every 500 samples based on this order, with the last group contains all the remaining  $781$ samples. Third, we plot the observed risk over the range of the blood mercury level in Figure ~\ref{fig8}. From the local pattern, we see that the overall decrease of the risk associated with the increase of the total blood mercury level. The curves of the estimated risks from the logistic Box-Cox model and the three misspecified logistic linear models over the range of the total blood mercury are also added in Figure ~\ref{fig8}, showing that the estimated risks of the square-root model are closest to those of the logistic Box-Cox model.  
 
\begin{figure}[t]
\centerline{\includegraphics[width=10cm, height=8cm]{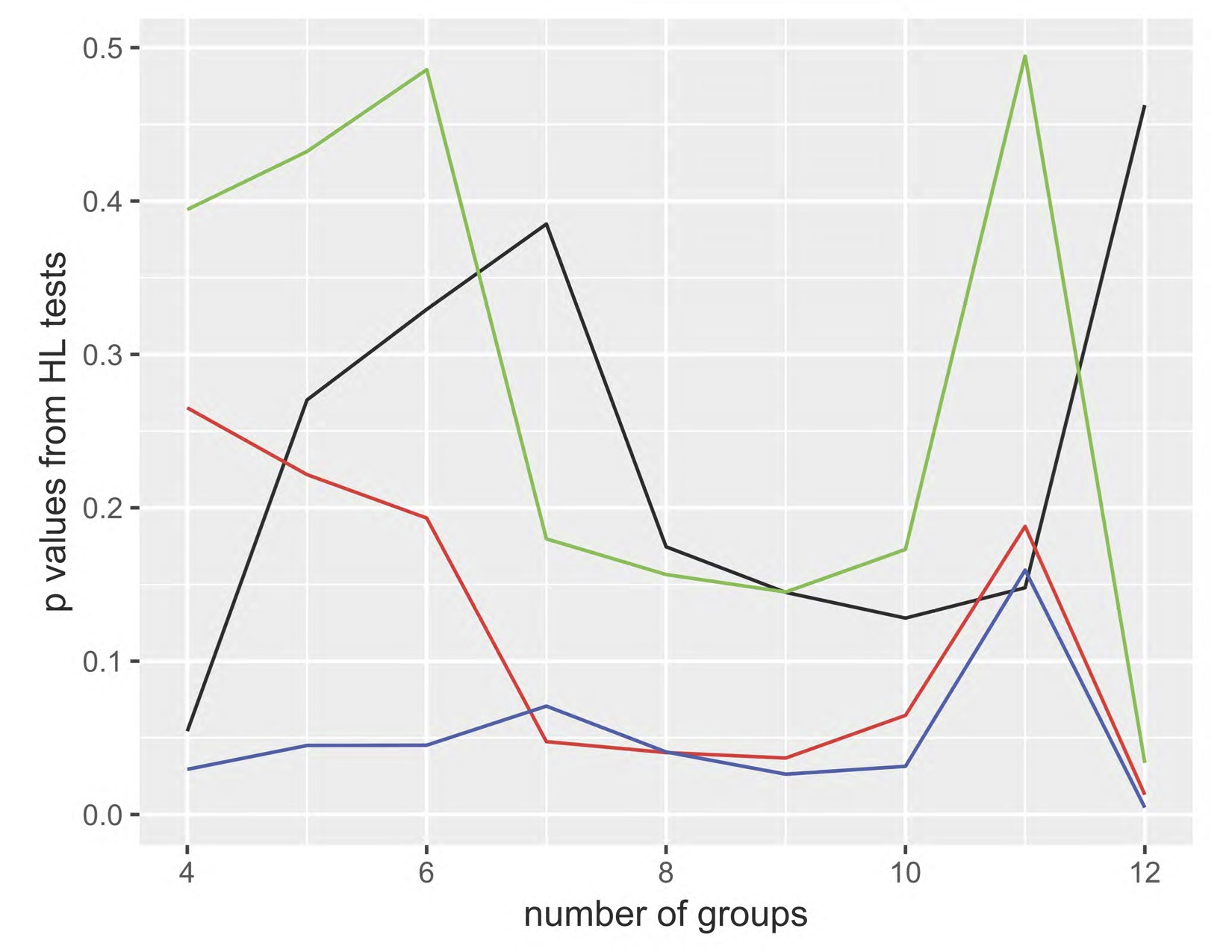}}
\caption{The p-value from the Hosemer and Lemeshow test versus the number of subgroups. The black segmented lines are from the fitted logistic Box-Cox  regression, the red segmented lines are from the linear logistic regression model on original scale, the green segment lines are the linear logistic regression model on square-root scale, and the blue segments are for the linear logistic regression model on logarithmic scale. }\label{fig9}
\end{figure}

 In addition to the graphical illustration, we compare the goodness of fit of the four models and also compare their predictions. We conduct the Hosemer and Lemeshow goodness of fit (GOF) test \cite{Hosmer2000} for the four models. This test statistic is the sum of the difference between the expected and the observed risks over pre-defined subgroups. To avoid the result depending on the number of subgroups, we vary the number of subgroups from $5$ to $12$ and for each partition of subgroups, we conduct the test. The resulting p-values are reported in Figure ~\ref{fig9}, which demonstrates that the square-root model is comparable to the Box-Cox model, while the logarithmic model is the worst in terms of goodness of fit.         

We also compare the predictions of the four models using 10-fold cross-validation, where we split the total samples equally into ten subgroups. Nine of the $10$ subgroups are combined as the training set that we use to fit the model, while the remaining one is the test set that we use for prediction based on the fitted model from the training set. When we iterate over all the possible combinations of nine subgroups,  the predicted risks from all the test sets become the prediction on all of the samples. We use r package \textit{caret}  \cite{JSSv028i05} for the data partitions, since its functions generate random samples within the level of the outcome and, therefore, the splits have the balanced class distributions. To compare the predictions, since all of the models have the same receiver operating characteristic (ROC) curve, we use two criteria, the mean absolute error and the mean squared error of the estimated risks. The mean absolute errors from the logistic Box-Cox model, the linear, the square-root and the log models are $0.1465, 0.1466, 0.1465,$ and $0.1466$, while their mean square errors are $0.0736, 0.0736, 0.0736,$ and $0.0737$. The errors from different models are close to each other, which is mainly due to the low exposure-disease association, ( the estimated risk ratio between the $95$th percentile and the $5$th percentile = $0.31$), (Refer to the (2, 2) panel of Figure ~\ref{fig1}). In summary, we conclude that the square-root model is comparable to the logistic Box-Cox model, and both outperform the log model. It is important to note that our analysis of NHANES data was not meant to illuminate association between mercury and depression, as it is most likely confounded to the degree that makes it impossible to argue that mercury protects against depression \cite{Ng:2013iz}.

\section{Discussion}

The logistic Box-Cox model is a formal method, which can accommodate the non-linear relationship between the log-odds and exposure via a shape parameter. The estimate of this parameter is determined based on the ML method. Particularly, we discuss the profile likelihood (PL) and the quasi-Newton methods. The profile likelihood can might lead to a local maximum solution.  The quasi-Newton method targets the global maximum, but it is sensitive to the initial point.  We recommend the PL method to provide the initial values for the quasi-Newton to guarantee a good starting point. In this way, we borrow strength from both methods in an attempt to obtain a superior overall approach.
              
As a non-linear model, the gradient of the log-odds with respect to the predictor is not constant. This encourages us to define the median effect, which represents the gradient over the entire distribution of the predictor. We generalize this quantity to the $X^{(q)}$ scale. In this way, we can compare it with the slope from the misspecified model based on the power transformation of $q$. The ARE is a measure of the distance between the large sample limiting value of the slope estimate and the median effect on the same scale. We see that even when model is misspecified, when there is little information, the slope estimated from the log transformation is can be close to the median effect relative to the magnitude of the median effect.           

We calculate the asymptotic standard deviation of the estimate of the shape parameter,  that of the estimated slope parameter given a certain scale, and that of the estimated median effect given a certain scale based on numerical methods. These quantities can help us design future studies. For instance, if we have prior knowledge about nonlinear relationship and skewed exposure, we can estimate the required sample size based on the desired accuracy for the estimate of the shape parameter. For the conducted studies with limited sample size, if disease is rare and association is not strong, the logarithm transformation provides stable measurement since now the more complex logistic Box-Cox model is less helpful due to the large estimated uncertainty on the parameter estimates. 

\section{Acknowledgements} 
This research is supported by NSERC through the Discovery Grants program, through the Canada Research Chair program, and through the NSERC Postdoctoral Fellowships Program and by the University of Victoria through a UVic Internal Research Grant and the UVic Faculty of Science.


\newpage
\begin{table}[ht]
\centering
\caption{The Simulation Settings} \label{tab:factors}
\begin{tabular}{l|l|l|l}
  \hline \hline
Distribution of $X$ & Shape of Relationship & Disease Rarity              & Exposure-Disease \\  
$\mbox{LN}(\mu, \sigma)$& $\lambda$      &$\mbox{P}( Y=1 | X = $5$th\mbox{Q})$& Association\\ \hline \hline
                                         & log $\lambda = 0$           &                                         &\\
   weakly skewed  $\sigma = 0.5$ & square-root $\lambda = 0.5$  & low $\mbox{P}_{1}= 0.02$  & weak $\mbox{R}_1=1.1$\\
   mild skewed $\sigma = 1$     & linear $\lambda = 1$         &                                             & mild  $\mbox{R}_2=2$\\
  highly skewed $\sigma = 2$  & square $\lambda = 2$        &  high $\mbox{P}_{2} = 0.1$   &  strong $\mbox{R}_3=5$  \\ \hline \hline
\end{tabular}
\end{table}

\begin{table}[ht]
\centering
\caption{ Estimates from the Misspecified Models, Estimated Median Effects, and ARE  } \label{tab:modelXq}
\begin{tabular}{l|l|l||c||l}
  \hline \hline
\multicolumn{3}{c||}{Logistic Linear Model}    & Logistic Box-Cox Model  & ARE \\ \cline{1-3}
$q$ value  & AIC & Slope (SE) & $\Delta^{*}_{q}$ & \\ 
\hline \hline
$q = 1$     & $5390$  & $-0.210 (0.035)$    & $-0.400$ & $0.475$\\ 
$q = 0.5$  &  $5381$ & $-0.307 (0.044)$    & $-0.322$ & $0.047$\\
$q = 0$     &  $5384$ & $-0.310 (0.042)$    & $-0.323$ & $0.040 $\\ \hline \hline
\end{tabular}
\end{table}

\bibliographystyle{abbrv}
\bibliography{DataTransformation}  

\textbf{Supplementary Material}

\textbf{Proof of Mathematical Theories and Calculations}

\noindent
\textbf{Proof of Proposition 1:} The log likelihood function is
	\begin{align*}
	l(\beta_{0}, \beta_{1}, \lambda | \bm Y,\bm X) = \sum_{i=1}^{n}y_{i}\left[\beta_{0} + \beta_{1} \left(\frac{x^{\lambda}_{i} - 1}{\lambda}\right)\right] - \log\left[1 + \exp\left(\beta_{0} + \beta_{1} \left(\frac{x^{\lambda}_{i} - 1}{\lambda}\right)\right)\right]
	\end{align*}

The score function is
\begin{equation*}
\left( \begin{array}{l}
\frac{\partial l}{\partial \beta_{0}}  \\
\frac{\partial l}{\partial \beta_{1}} \\
\frac{\partial l}{\partial \lambda}  \end{array} \right) = \left( \begin{array}{l}
\sum\limits_{i = 1}^{n}(y_i - p_i)  \\
\sum\limits_{i = 1}^{n}(y_i - p_i)\nu_{i}  \\
\sum\limits_{i = 1}^{n}(y_i - p_i)\beta_{1}\left(\frac{x^{\lambda}_{i}\ln{x_{i}} - \nu_{i}}{\lambda}\right) \end{array} \right), 
\end{equation*}

and the Hessian  matrix is 
   \begin{align*}
   H &= -
   \begin{bmatrix}
   \sum\limits_{i = 1}^{n}p_{i}q_{i}, &\sum\limits_{i = 1}^{n}p_{i}q_{i}\nu_{i},  &\sum\limits_{i = 1}^{n}p_{i}q_{i}\beta_{1}\frac{\partial \nu_{i}}{\partial \lambda} \\ 
   \sum\limits_{i = 1}^{n}p_{i}q_{i}\nu_{i}, &\sum\limits_{i = 1}^{n}p_{i}q_{i}\nu^2_{i},  &\sum\limits_{i = 1}^{n}p_{i}q_{i}\beta_{1}\nu_{i}\frac{\partial \nu_{i}}{\partial \lambda} - (y_{i} - p_{i})\frac{\partial \nu_i}{\partial \lambda} \\ 
   \sum\limits_{i = 1}^{n}p_{i}q_{i}\beta_{1}\frac{\partial \nu_{i}}{\partial \lambda}, &\sum\limits_{i = 1}^{n}p_{i}q_{i}\beta_{1}\nu_{i}\frac{\partial \nu_i}{\partial \lambda} - (y_{i} - p_{i})\frac{\partial \nu_i}{\partial \lambda},  &\sum\limits_{i = 1}^{n}p_{i}q_{i}\beta^{2}_{1}\left(\frac{\partial \nu_{i}}{\partial \lambda}\right)^{2} - (y_{i} - p_{i})\beta_{1}\frac{\partial^2 \nu_{i}}{\partial \lambda^2}
   \end{bmatrix}, 
      \end{align*}
where $\nu_{i} = \frac{x^{\lambda}_{i} - 1}{\lambda}$, and $q_{i}=(1-p_{i})$.

 We prove the leading principal minors of the Hessian  matrix are all negative.   
 
 \begin{align*}  
   \det{H_{11}} &= -\sum_{i = 1}^{n}p_{i}q_{i} < 0 \\
   \det{(H_{22})} &= -\det{\left(\begin{bmatrix}
   \sum_{i = 1}^{n}p_{i}q_{i}, &\sum_{i = 1}^{n}p_{i}q_{i}\nu_{i} \\ 
   \sum_{i = 1}^{n}p_{i}q_{i}\nu_{i}, &\sum_{i = 1}^{n}p_{i}q_{i}\nu^2_{i} 
   \end{bmatrix}\right)} = -\sum_{i \neq j}p_{i}q_{i}p_{j}q_{j}(\nu_{i} - \nu_{j})^2 < 0
  \end{align*} 
 
As for $\det{(H_{33})}$, we can write $H$ as a sum of matrices, $\{H^{i}\}_{i=1}^{n}$, and each of $H^{i}$  are negative semi-definite, particular 
\begin{equation*}
H^{i} = \begin{bmatrix}
    p_{i}q_{i}, & p_{i}q_{i}\nu_{i},  & p_{i}q_{i}\beta_{1}\frac{\partial \nu_{i}}{\partial \lambda} \\ 
    p_{i}q_{i}\nu_{i}, & p_{i}q_{i}\nu^2_{i},  & p_{i}q_{i}\beta_{1}\nu_{i}\frac{\partial \nu_{i}}{\partial \lambda} - (y_{i} - p_{i})\frac{\partial \nu_i}{\partial \lambda} \\ 
    p_{i}q_{i}\beta_{1}\frac{\partial \nu_{i}}{\partial \lambda}, &p_{i}q_{i}\beta_{1}\nu_{i}\frac{\partial \nu_i}{\partial \lambda} - (y_{i} - p_{i})\frac{\partial \nu_i}{\partial \lambda},  & p_{i}(1-p_{i})\beta^{2}_{1}\left(\frac{\partial \nu_{i}}{\partial \lambda}\right)^{2}- (y_{i} - p_{i})\beta_{1}\frac{\partial^2 \nu_{i}}{\partial \lambda^2}
   \end{bmatrix}
\end{equation*}
We have
       \begin{align*}
       \det{(H^{i}_{1 \times 1})} & = - p_{i}q_{i} \leq 0\\
       \det{(H^{i}_{2 \times 2})} & =  0 \\
       \det{(H^{i}_{3 \times 3})} & = -\frac{1}{\lambda^{4}}p_{i}(1-p_{i})(y_{i}-p_{i})^{2}\left(1 + x^{\lambda}_{i}(-1+\lambda \mbox{Ln}(x_{i}))\right)^{2} \leq 0. 
   \end{align*}	
Based on Minkowski Determinant Theorem, we have $\det{(H)} \leq \sum_{i=1}^{n}\det{(H^{i})}$ and then $\det{(H)} \neq 0$.  Therefore, the Hessian  matrix is negative definite.

\noindent
\textbf{Proof of the Proposition 2:} 
Let $Z \sim \mbox{N}(0,1)$ and then we have $X= \mbox{exp}\left(\mu + Z\sigma \right)$. Also denote $V = \frac{X^{\lambda} - 1}{\lambda}$.  
For a single observation, we can calculate the Fisher information matrix as follows. 
{\small
\begin{eqnarray} 
\nonumber I_{1}\left(\begin{array}{c}
\beta_{0} \\
\beta_{1} \\
\lambda
\end{array}\right) &=& - \mbox{E}\left[ 
   \begin{array}{ccc}
          \frac{\partial^{2} l(X, Y)}{\partial \beta_{0}^{2}} & \frac{\partial^{2} l(X, Y)}{\partial \beta_{1} \partial \beta_{0}} & \frac{\partial^{2} l(X, Y)}{\partial \lambda \partial \beta_{0}} \\
          \frac{\partial^{2} l(X, Y)}{\partial \beta_{1} \partial \beta_{0}} & \frac{\partial^{2} l(X, Y)}{\partial \beta_{1}^{2}}  &  \frac{\partial^{2} l(X, Y)}{\partial \lambda \partial \beta_{1}} \\
          \frac{\partial^{2} l(X, Y)}{\partial \lambda \partial \beta_{0}} & \frac{\partial^{2} l(X, Y)}{\partial \lambda \partial \beta_{1}} & \frac{\partial^{2} l(X, Y)}{\partial \lambda^{2}}  
   \end{array}\right] 
=  - \mbox{E}\left[ 
   \begin{array}{ccc}
          \mbox{E}\left( \frac{\partial^{2} l(X, Y)}{\partial \beta_{0}^{2}} | X \right) & \mbox{E}\left(\frac{\partial^{2} l(X, Y) }{\partial \beta_{1} \partial \beta_{0}} | X \right) & \mbox{E}\left( \frac{\partial^{2} l(X, Y)}{\partial \lambda \partial \beta_{0}} | X \right ) \\
          \mbox{E}\left( \frac{\partial^{2} l(X, Y)}{\partial \beta_{1} \partial \beta_{0}} | X \right) & \mbox{E}\left( \frac{\partial^{2} l(X, Y)}{\partial \beta_{1}^{2}} | X \right) &  \mbox{E}\left( \frac{\partial^{2} l(X, Y)}{\partial \lambda \partial \beta_{1}} | X \right)\\
          \mbox{E}\left( \frac{\partial^{2} l(X, Y)}{\partial \lambda \partial \beta_{0}} | X \right) & \mbox{E}\left( \frac{\partial^{2} l(X, Y)}{\partial \lambda \partial \beta_{1}} | X \right) & \mbox{E} \left(\frac{\partial^{2} l(X, Y)}{\partial \lambda^{2}} | X \right)  
   \end{array}\right]\\
\label{eq:fisherinfo}  &=& \left[ 
      \begin{array}{ccc}
      \int_{-\infty}^{+\infty}  pq \phi(z) \mbox{d}z & \int_{-\infty}^{+\infty}   p q v\phi(z) \mbox{d}z     & \int_{-\infty}^{+\infty}  pq \beta_{1} \frac{\partial v}{\partial \lambda} \phi(z) \mbox{d} z \\
      \int_{-\infty}^{+\infty}  pqv \phi(z) \mbox{d}z & \int_{-\infty}^{+\infty}   pqv^{2}\phi(z) \mbox{d}z     & \int_{-\infty}^{+\infty} pq\beta_{1}v \frac{\partial v}{\partial \lambda} \phi(z) \mbox{d}z \\
       \int_{-\infty}^{+\infty} pq \beta_{1} \frac{\partial v}{\partial \lambda} \phi(z) \mbox{d}z & \int_{-\infty}^{+\infty} pq \beta_{1} v \frac{\partial v}{\partial \lambda} \phi(z) \mbox{d}z     & \int_{-\infty}^{+\infty} pq \left(\beta_{1}\frac{\partial v}{\partial \lambda} \right)^{2} \phi(z) \mbox{d}z 
     \end{array} \right],        
\end{eqnarray} 
where $\phi(z)$ is the probability density function of the standard normal distribution. And, denote $\mbox{C}_{33}$ as the matrix cofactor of $I_{1}$. We have  
\begin{equation} \label{eq:avarequation}
\mbox{Avar}(\hat{\lambda}) = \frac{1}{\mbox{det}[I_{1}]} \mbox{C}_{33}.
\end{equation}}
Since 
\begin{equation*}
\lim_{\beta_{1} \to 0} p(x) = \frac{\mbox{exp}(\beta_{0})}{1 + \mbox{exp}(\beta_{0})},
\end{equation*}
based on Lebesgue's Dominated Convergence Theorem \cite{Bartle1995}, we have 
\begin{equation}
\lim_{\beta_{1} \to 0} \mbox{C}_{33} = \left[\frac{\mbox{exp}(\beta_{0})}{\left(1 + \mbox{exp}(\beta_{0})\right)^{2}}\right]^{2}\frac{\mbox{exp}(2 \lambda\mu) \left[\mbox{exp}(\lambda^2 \sigma^2)-\mbox{exp}(\frac{\lambda^2 \sigma^2}{2})\right]}{\lambda^2},
\end{equation}
and
\begin{equation}
\lim_{\beta_{1} \to 0} \beta_{1}^{2}\mbox{det}(I_{1}(\bm \theta )) =  \left[\frac{\mbox{exp}(\beta_{0})}{\left(1 + \mbox{exp}(\beta_{0})\right)^{2}}\right]^{3} \frac{\mbox{exp}\left(\frac{3 \lambda^2 \sigma^2 + 8 \lambda \mu}{2}\right)\left(\mbox{exp}\left(\frac{\lambda^2 \sigma^2}{2}\right) -\frac{\lambda^2 \sigma^2}{2}-1\right) \sigma^2 }{2 \lambda^4} 
\end{equation}
Therefore, we demonstrate that $\mbox{Avar}(\hat{\lambda}) = \mbox{O}(\beta_{1}^{-2}).$

\textbf{Numerical calculation of $\mbox{Avar}(\hat{\lambda})$.}
In general, we cannot get the closed form expression for the integrals shown in equation (\ref{eq:avarequation}). However, numerical evaluation of the integrals is possible by using the
Gaussian-Hermite Quadrature (GHQ). In GHQ, we use the following approximation.  
\begin{equation*}  
 \int_{-\infty}^{+\infty} e^{-t^2} f(t)\,dt \approx \sum_{i=1}^m w_i f(t_i)  
\end{equation*}
where $m$ is the number of sample points used, and $\{t_{i}\}_{(i = 1, \cdots, m)}$ are the roots of the Hermite polynomial $H_{m}(t) (i = 1, 2, ..., m)$, and the associated weights $w_{i}$ are given by \cite{Abramowitz1972}
\begin{equation*} 
    w_i = \frac {2^{m-1} m! \sqrt{\pi}} {m^2[H_{m-1}(x_i)]^2}.
\end{equation*}

\textbf{Numerical Calculation of $\mbox{Avar}(\hat{\bm \gamma}_{q})$}

We calculate the asymptotic variance of $\hat{\bm \gamma}_{q}$ for a data with a single observation as follows.  
First, the gradient of the likelihood is 
\begin{equation*}
\nabla l_{1} (\bm \gamma_{q})= \left( \begin{array}{l}
                               \frac{\partial l}{\partial \gamma_{0q}} \\
                               \frac{\partial l}{\partial \gamma_{1q}} 
\end{array}\right) =  \left( \begin{array}{l}
                               Y - P^{*} \\
                               (Y - P^{*})W_{q}   
\end{array}\right),
\end{equation*}
where $P^{*} = \frac{\mbox{exp}(\gamma_{0q} + \gamma_{1q}W_{q})}{ 1 + \mbox{exp}(\gamma_{0q} + \gamma_{1q}W_{q})}.$
Then, its variance is 
\begin{eqnarray*}
V_{1}(\bm \gamma_{q}) &=& \mbox{Var}(\nabla l_{1}(\bm \gamma_{q}))\\
                                &=& \mbox{E}\left( \nabla l_{1}(\bm \gamma_{q}) \left( \nabla l_{1}(\bm \gamma_{q}) \right)^{T} \right)\\
                                &=& \left( \begin{array}{ll}
  \mbox{E}\left[(Y-P^{*})^{2}\right]  & \mbox{E}\left[(Y-P^{*})^{2}W_{q}\right]\\
 \mbox{E}\left[(Y-P^{*})^{2}W_{q}\right] & \mbox{E}\left[(Y-P^{*})^{2}W^{2}_{q}\right] 
                                \end{array}\right)\\
                                 &=& \left( \begin{array}{ll}
  \mbox{E}\left[ P - 2PP^{*} + P^{*2}\right]  & \mbox{E}\left[(P - 2PP^{*} + P^{*2})W_{q}\right]\\
  \mbox{E}\left[(P - 2PP^{*} + P^{*2})W_{q}\right] & \mbox{E}\left[(P - 2PP^{*} + P^{*2})W^{2}_{q}\right] 
                                \end{array}\right).  
\end{eqnarray*}
Second, the Hessian matrix of the likelihood is  
\begin{equation*}
H(l_{1}) = \left( \begin{array}{ll}
  \frac{\partial^{2} l}{\partial \gamma^{2}_{0q}} &  \frac{\partial^{2} l}{\partial \gamma_{0q} \partial \gamma_{1q}}\\
  \frac{\partial^{2} l}{\partial \gamma_{0q} \partial \gamma_{1q}} & \frac{\partial^{2} l}{\partial \gamma^{2}_{1q}} 
\end{array}\right) =  \left( \begin{array}{ll}
                                -P^{*}(1 - P^{*})   & -P^{*}(1 - P^{*})W_{q} \\
                                -P^{*}(1 - P^{*})W_{q} & -P^{*}(1 - P^{*})W_{q}^{2}. 
\end{array}\right)
\end{equation*}
with the expectation of 
\begin{equation*}
J_{1}(\bm \gamma_{q}) = -\mbox{E}(H(l_{1})) = \left( \begin{array}{ll}
                                \mbox{E}(P^{*}(1 - P^{*}))   & \mbox{E}(P^{*}(1 - P^{*})W_{q}) \\
                                \mbox{E}(P^{*}(1 - P^{*})W) & \mbox{E}(P^{*}(1 - P^{*})W^{2}_{q})
\end{array} \right).
\end{equation*}

We need to solve equation (\ref{eq:mismodel}) in the manuscript to get the large-sample limiting coefficient estimate of $\bm \gamma_{q}$ and then to calculate equation (\ref{eq:AvarGamma}) in the manuscript to get the asymptotic standard deviations of the MLE of $\bm \gamma_{q}$'s. Since we can not get the closed form solution for both of them, we do both based on numerical calculation. First, given $\mu$ and $\sigma^{2}$, we simulate $N$ samples from the distribution of $X$, and then get the corresponding $W_{q}$. We use the sample mean to approximate the expectation. That is, for equation (\ref{eq:mismodel}) in the manuscript, we solve
\begin{equation} \label{eq:average}
\frac{1}{N}\sum\limits_{i = 1}^{N} \left[\left(\begin{array}{c}
                               1  \\
                               w_{qi}
                      \end{array}\right)\left(p_{i} - p^{*}_{i}\right)\right] = \bm 0,
\end{equation}   
where \begin{equation*}
 p_{i} = \mbox{E}( Y | X = x_{i}) = \frac{\mbox{exp}\left(\beta_{0} + \beta_{1}\frac{x^{\lambda}_{i} - 1}{\lambda}\right)}{ 1 + \mbox{exp}\left(\beta_{0} + \beta_{1}\frac{x^{\lambda}_{i} - 1}{\lambda}\right)},
\end{equation*}
and 
\begin{equation*}
p^{*}_{i} = \frac{ \mbox{exp}(\gamma_{0q} + \gamma_{1q}w_{qi}) }{ 1 +  \mbox{exp}(\gamma_{0q} + \gamma_{1q}w_{qi}) }.
\end{equation*}
If we treat $(w_{qi}, p_{i})_{\left( i = 1, \cdots, N \right)}$ or say $(x_{i}, q, p_{i})_{\left( i = 1, \cdots, N \right)}$ as our data, then the equation (\ref{eq:average}) can be solved by standard logistic regression, which allows us to pretend that the outcome is not binary. We call the standard errors of the coefficients obtained in this manner as the simulation errors, since these errors are introduced by the limitation of a finite simulated sample size.  And for equation (\ref{eq:AvarGamma}) in the manuscript, we get 
\begin{equation} 
\mbox{Avar}(\hat{\bm \gamma}_{q}) \approx \hat{J}^{-1}_{1}(\bm \gamma_{q}^{*})\hat{V}_{1}(\bm \gamma_{q}^{*})\hat{J}^{-1}_{1}(\bm \gamma_{q}^{*}),
\end{equation}
where 
\begin{equation*}
\hat{J}_{1}(\bm \gamma_{q}^{*}) = \left( \begin{array}{ll}
\frac{1}{N}\sum\limits_{i = 1}^{N}\left[p^{**}_{i}(1 - p^{**}_{i})\right]   &  \frac{1}{N}\sum\limits_{i = 1}^{N}\left[p^{**}_{i}(1 - p^{**}_{i})w_{qi}\right] \\
\frac{1}{N}\sum\limits_{i = 1}^{N}\left[p^{**}_{i}(1 - p^{**}_{i})w_{qi}\right] &  \frac{1}{N}\sum\limits_{i = 1}^{N}\left[p^{**}_{i}(1 - p^{**}_{i})w^{2}_{qi}\right]
\end{array} \right),
\end{equation*}
and
\begin{equation*}
\hat{V}_{1}(\bm \gamma^{*}_{q}) = 
\left( \begin{array}{ll}
\frac{1}{N}\sum\limits_{i = 1}^{N}\left[ p_{i} - 2p_{i}p^{**}_{i} + p^{**2}_{i}\right]  &  \frac{1}{N}\sum\limits_{i = 1}^{N}\left[(p_{i} - 2p_{i}p^{**}_{i} + p^{**2}_{i})w_{qi}\right]\\
\frac{1}{N}\sum\limits_{i = 1}^{N}\left[(p_{i} - 2p_{i}p^{**}_{i} + p^{**2}_{i})w_{qi}\right] &\frac{1}{N}\sum\limits_{i = 1}^{N}\left[(p_{i} - 2p_{i}p^{**}_{i} + p^{**2}_{i})w^{2}_{qi}\right] 
\end{array}\right),
\end{equation*}
where $\bm \gamma^{*}_{q}$ is the solution of equations (\ref{eq:average}) and
\begin{equation*}
p^{**}_{i} = \frac{ \mbox{exp}(\gamma^{*}_{0q} + \gamma^{*}_{1q}w_{qi}) }{ 1 +  \mbox{exp}(\gamma^{*}_{0q} + \gamma^{*}_{1q}w_{qi}) }.
\end{equation*}

In our example, we choose $N = 100,000.$

\textbf{Additional Figures}

\begin{figure}[h!btp]
\begin{center}
\vspace{ 0.2in}
\includegraphics[scale = 0.3 ]{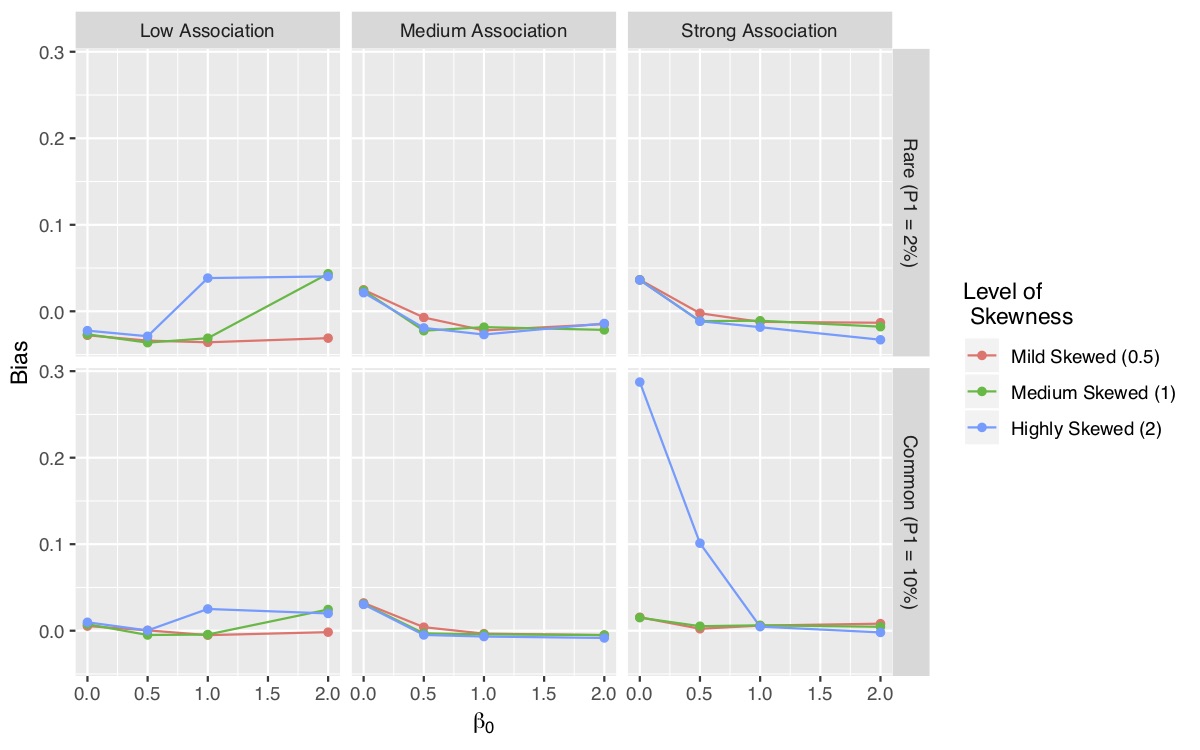}
\caption{The bias of $\hat{\beta}_{0}$ in the logistic Box-Cox model based on the BFGS algorithm under $72$ settings.} \label{fig:beta0bias}
\end{center}
\end{figure}

\begin{figure}[h!btp]
\begin{center}
\vspace{ 0.2in}
\includegraphics[scale = 0.3]{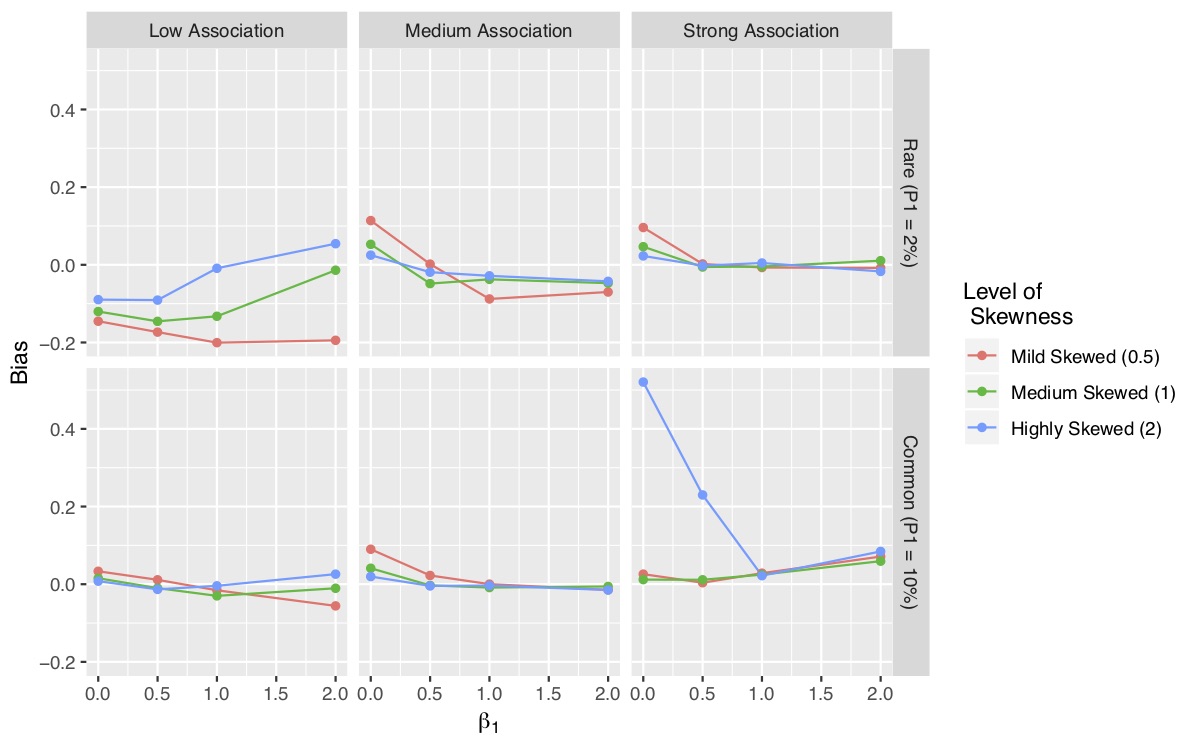}
\caption{The bias of $\hat{\beta}_{1}$ in the logistic Box-Cox model based on the BFGS algorithm under $72$ settings.} \label{fig:beta1bias}
\end{center}
\end{figure}

\begin{figure}[h!btp]
\begin{center}
\vspace{ 0.2in}
\includegraphics[scale = 0.3 ]{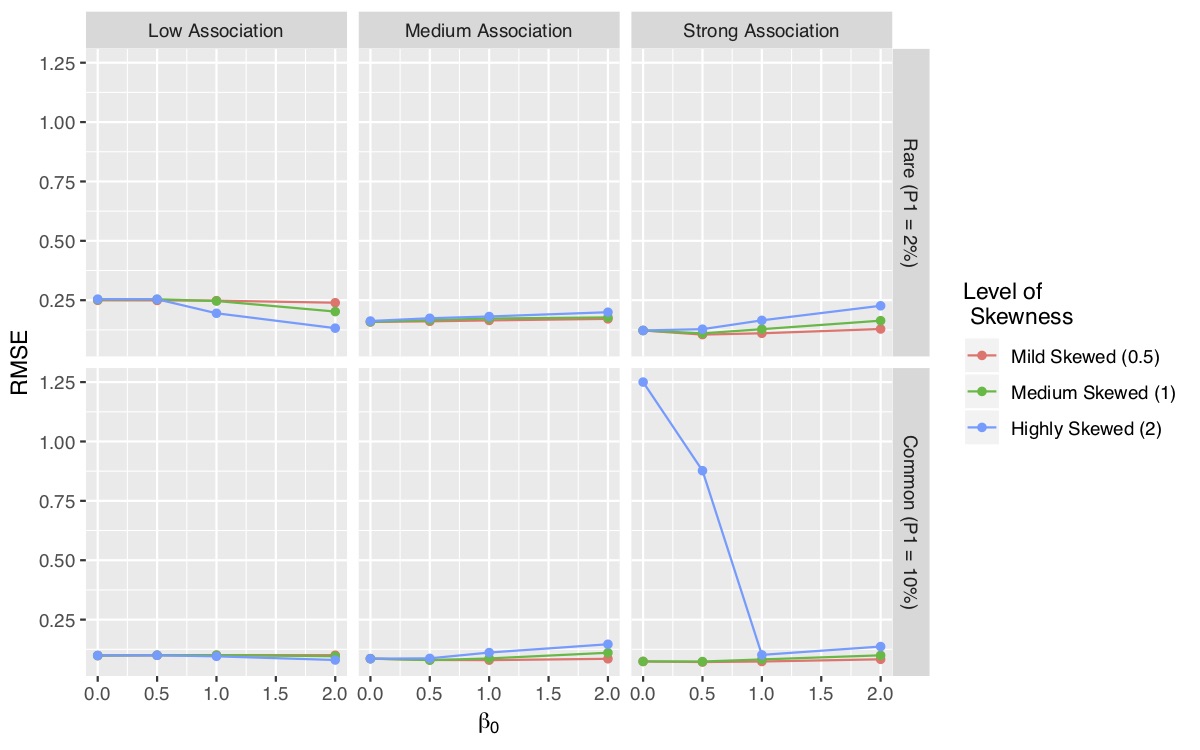}
\caption{The RMSE of $\hat{\beta}_{0}$ in the logistic Box-Cox model based on the BFGS algorithm under $72$ settings.} \label{fig:beta0mse}
\end{center}
\end{figure}

\begin{figure}[h!btp]
\begin{center}
\vspace{ 0.2in}
\includegraphics[scale = 0.3 ]{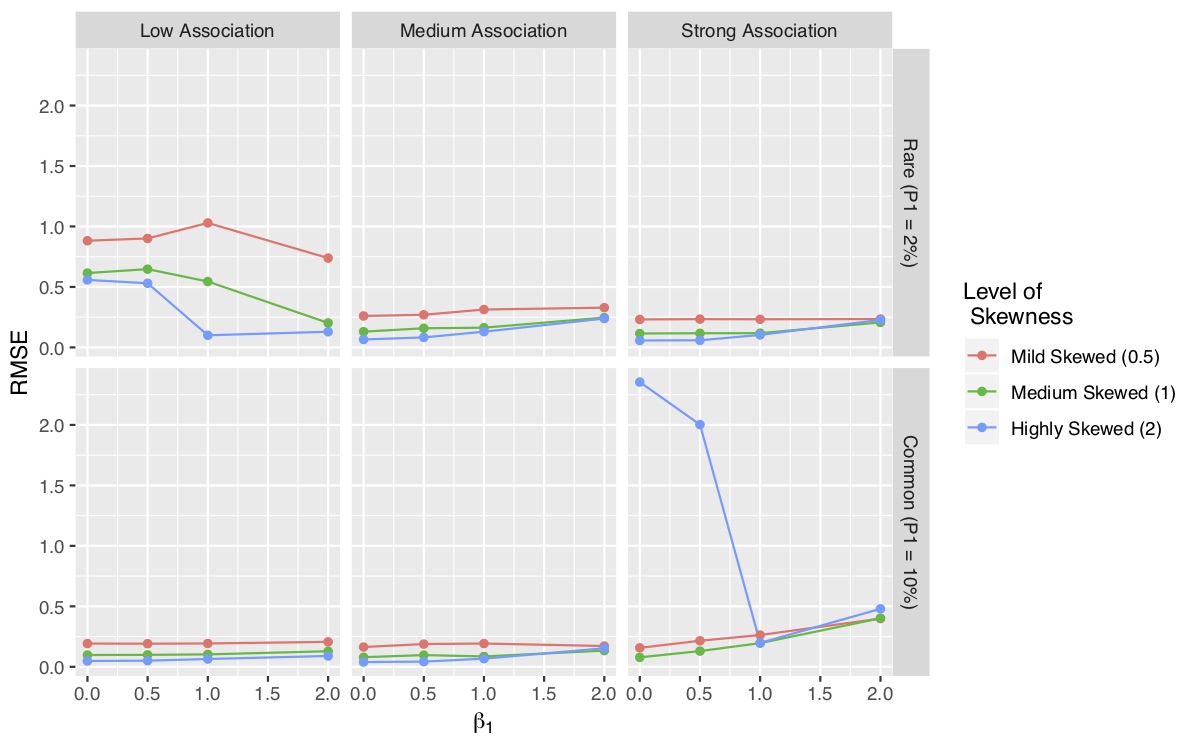}
\caption{The RMSE of $\hat{\beta}_{1}$ in the logistic Box-Cox model based on the BFGS algorithm under $72$ settings.} \label{fig:beta1MSE}
\end{center}
\end{figure}

\begin{figure}[h!btp]
\begin{center}
\vspace{0.2in}
\includegraphics[scale=0.3]{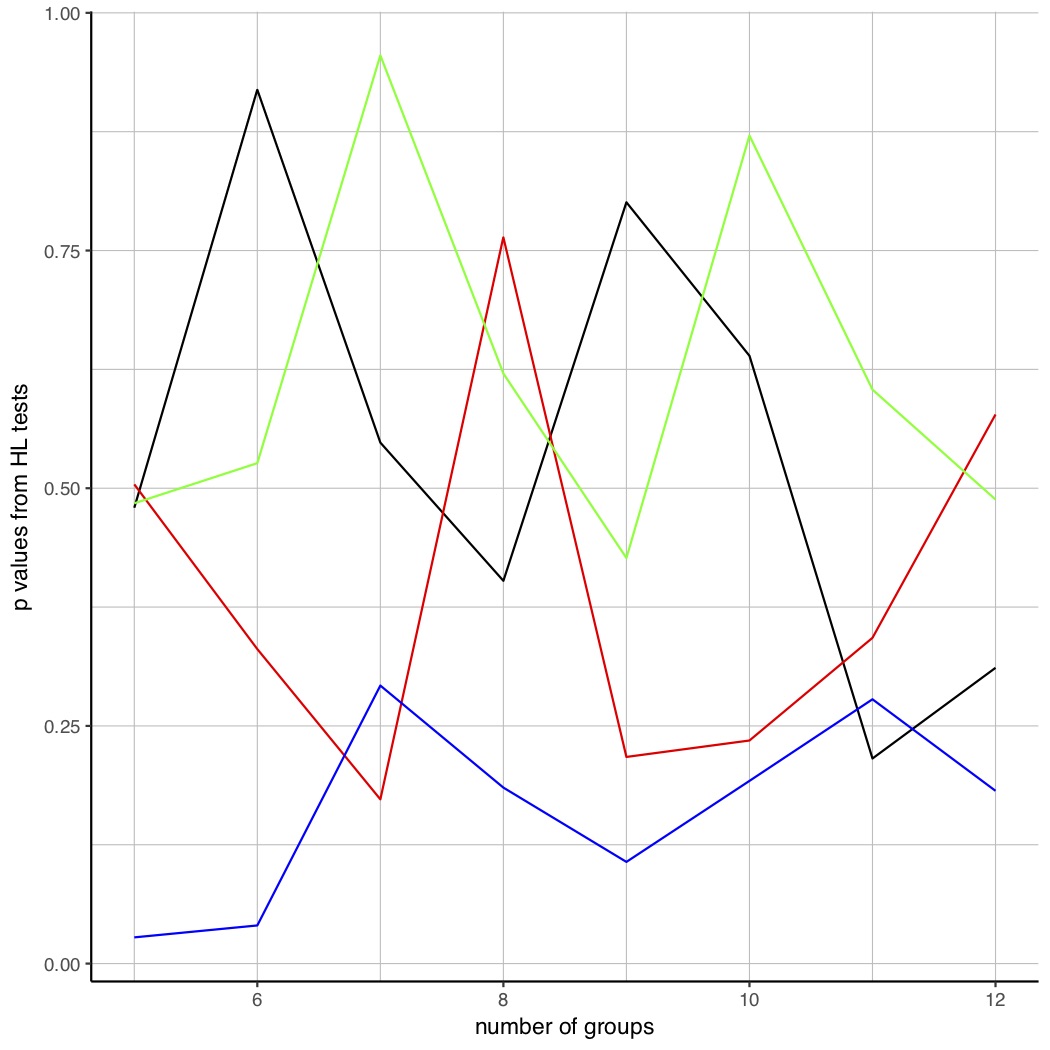}
\caption{P value from the Hosemer and Lemeshow test versus the number of groups. The black segments are for the fitted logistic Box-Cox  regression, the red segments are for the linear logistic regression model on original scale, the green segments are the linear logistic regression model on square-root scale, and the blue segments are for the linear logistic regression model on logarithmic scale} \label{fig:goodp}
\end{center}
\end{figure}

%
%
%
%
%
\end{document}